\documentclass[epj,nopacs]{svjour}
\usepackage{graphics}

\newcommand{\be}{\begin{equation}}
\newcommand{\ee}{\end{equation}}

\begin{document}

\thispagestyle{empty}

\title{Thermal correction to the Casimir force,
radiative heat transfer, and an experiment}

\author{V.~B.~Bezerra\inst{1} \and
G.~Bimonte\inst{2,3} \and
G.~L.~Klimchitskaya\inst{4,}\thanks{On leave from North-West 
Technical University,
{\protect \\} Millionnaya St. 5, St.Petersburg,
191065, Russia}  \and
V.~M.~Mostepanenko\inst{4,}\thanks{On leave from Noncommercial
Partnership ``Scientific {\protect \\} Instruments'', Tverskaya St. 11, Moscow,
103905, Russia} \and
C.~Romero\inst{1}
}

\institute{Department of Physics, Federal University of Para\'{\i}ba,
C.P.5008, CEP 58059--970, Jo\~{a}o Pessoa, Pb-Brazil \and
Dipartimento
di Scienze Fisiche, Universit\`{a}
di Napoli Federico II, Complesso
Universitario MSA, {\protect \\} Via Cintia
I---80126 Napoli, Italy \and
INFN, Sezione di
Napoli, Napoli, Italy \and
Center of Theoretical Studies and Institute for Theoretical Physics,
Leipzig University, {\protect \\}
D-04009, Leipzig, Germany 
}
\date{Received: date / Revised version: date}
%
\abstract{
The low-temperature asymptotic expressions for  the
Casimir interaction between two real metals described 
by Leontovich surface impedance are obtained in the 
framework of thermal quantum field theory. It is shown
that the Casimir entropy computed using the impedance 
of infrared optics vanishes in the limit of zero 
temperature. By contrast, the  Casimir entropy computed 
using the impedance of the Drude model attains at zero 
temperature a positive value which depends on the 
parameters of a system, i.e., the Nernst heat theorem 
is violated. Thus, the impedance of infrared optics 
withstands the thermodynamic test, whereas the impedance
of the Drude model does not. We also perform a 
phenomenological analysis of the thermal Casimir force 
and of the radiative heat transfer through a vacuum 
gap between real metal plates. The characterization 
of a metal by means of the Leontovich impedance of 
the Drude model is shown to be inconsistent with
experiment at separations of a few hundred nanometers.
A modification of the impedance of infrared optics is 
suggested taking into account relaxation processes. 
The power of  radiative heat transfer predicted from 
this impedance is several times less than previous 
predictions due to different contributions from the
transverse electric evanescent waves.  The physical 
meaning of low frequencies in the Lifshitz formula 
is discussed. It is concluded that new measurements 
of radiative heat transfer are required to find out 
the adequate description of a metal in the theory of 
electromagnetic fluctuations.
}
\authorrunning{V.~B.~Bezerra et al.}
\titlerunning{Thermal correction to the Casimir force and
radiative heat transfer}

\maketitle
\section{Introduction}
\label{intro}
During the last few years, complicated problems connected
with the
concept of quantum fluctuations generated much interest
among
specialists in gravitation and cosmology, 
dispersion forces, Bose-Einstein
condensation,
nanotechnology, radiative heat transfer and related subjects.
Van der Waals
and Casimir forces, which are different kinds
of dispersion forces,
arise from zero-point oscillations of the
electromagnetic field and thermal
photons. They act between closely
spaced macrobodies, between a
microparticle and a macrobody or
between two microparticles. The theory of
dispersion forces is based on
quantum statistical physics. For real bodies
at temperature $T$ described
by  a dielectric permittivity depending
only on frequency, the
van der Waals and Casimir forces acting between
them are calculated in
the framework of Lifshitz theory
\cite{Lifshitz,DLP,LP-SPII}.
Originally  Lifshitz theory was developed using the concept
of an
oscillating electromagnetic field and the
fluctuation-dissipation
theorem. Later the main equations of this theory, including the
famous
Lifshitz formula, were rederived in different
formalisms
\cite{4,5,6,7} and in particular on the basis of thermal quantum field
theory in the
Matsubara formulation \cite{8}. The Lifshitz theory was
recently used
for the interpretation of many experiments on the
measurement of the
Casimir force
\cite{Lam97,12,13,14,15,16,17,18,Decca1,Decca2,Decca3,D4,20,20a,20aa,20b},
in
the application of the
Casimir and van der Waals forces in nanotechnology
\cite{21,22,23,23a,23aa},
in Bose-Einstein condensation \cite{24,25} and also
for the
description of   radiative heat transfer between   two bodies
at
different temperatures through a vacuum gap \cite{27,BiPRL}.

The
application of   Lifshitz theory to real metallic bodies at
nonzero
temperature has led to   controversial results, depending
on the used model of
dielectric permittivity. If the boundary bodies
are described by the
free electron plasma model the ensuing
thermal correction to the Casimir
force \cite{28,29} is in qualitative
agreement with that obtained for
ideal metals on the basis of thermal
quantum field theory in Matsubara
formulation \cite{8,30}.
If, however, metal boundaries are described by
the Drude model
which takes relaxation into account, the thermal
correction at
short separations is many hundred times larger than for ideal
metals
and two times smaller than the latter at large
separations
\cite{BS_00,32}.
In the case of perfect crystal lattices with
no impurities, the entropy of a fluctuating field (i.e., the
Casimir entropy) calculated using the Drude model
takes a negative value when the temperature
vanishes, i.e., the Nernst heat theorem is violated \cite{36,37}.
In \cite{PB26,BrPRE05} it was argued that for real metals with
impurities this violation does not occur, but in the case of
perfect crystal lattices the problem remains unsolved. If
the dielectric permittivity of the plasma model is chosen
the Nernst heat theorem is satisfied \cite{36,37}.
Importantly, the
application of the Drude dielectric function
was  found
to be inconsistent
with experiment \cite{18,Decca1,Decca2,Decca3,D4}. On the
contrary, the plasma
model approach is consistent with  experimental data \cite{modPl}.

Interestingly, similar problem arises in the theory of the thermal
Casimir force acting between dielectrics. If the static dielectric
permittivity of dielectric materials is supposed to be finite,
Lifshitz theory is found \cite{PB31,PB32} to be in agreement  with
thermodynamics. If, however, the dc conductivity of dielectric
materials is taken into account, the Nernst heat theorem for the
entropy of a fluctuating field is violated \cite{PB31,PB32}. The same
is true in the metal-dielectric configuration \cite{PB33}
depending on the finiteness of the static permittivity of a dielectric
plate \cite{PB34,PB34a}. Thus, thermodynamics provides a test for the validity
of various models of material properties: only the
thermodynamically-consistent models should be used.
In this connection it is notable that just the Drude model of a metal,
which was shown to imply a
violation of the Nernst heat theorem for the Casimir entropy, was
found to be in contradiction with experiments
(controversial opinions on this subject
can be found in \cite{PB37,statusJPA06}).

As an alternative to the dielectric model, the optical properties of
a metal surface can be characterized in terms of
the Leontovich surface impedance, together with the corresponding
boundary conditions \cite{LLP-ED}.
In the framework of the Lifshitz theory it
was employed in \cite{Kats} at zero temperature and in
\cite{PB41,34} at nonzero temperature.
It should be kept in mind that both models of real
metals, the one based on the frequency-dependent dielectric permittivity
as well as the one using Leontovich impedance, are approximations,
each having its own range of validity. The concept of
a frequency-dependent permittivity is inapplicable in the
frequency region of the anomalous skin effect, where spatial
dispersion contributes critically. Physically this is explained
by the fact that for these frequencies the penetration depth of the
electromagnetic field inside a metal becomes of the same order
as the mean free path of conduction electrons and remains much
less than the distance traveled by an electron during the
period of the field. As a result, the
spatial non-uniformity of the field renders impossible a macroscopic
description in terms of a dielectric permittivity depending only
on the frequency \cite{LLP-ED}. On the other hand, impedance
boundary conditions retain their validity
at the frequencies of the anomalous skin effect because the
field inside a metal near the surface can be considered as a plane
wave propagating perpendicular to the surface.

However, Leontovich impedance, though
applicable to the description of the anomalous skin effect, cannot be
used at short separations where its magnitude is not much less than
unity. The physical reason for this is that with decreasing
separations the relevant characteristic frequencies enter the optical
region, where the field inside a metal cannot be considered anymore
as propagating perpendicularly to the metal surface. In this frequency
region the concept of dielectric permittivity depending only on the
frequency is satisfactory (because the distance traveled by an
electron during the period of the field is
much less than the penetration depth). However, the magnitude of
the dielectric permittivity is not large enough, and, as a consequence,
the angle of refraction depends on the angle of incidence.
Thus, the impedance boundary condition does not apply.

The explicit analytic forms of the impedance function are
available in the asymptotic regions of the normal and anomalous skin
effect and in infrared optics. As was noticed without a detailed
proof in \cite{34}, the Casimir entropy calculated using the
impedance of infrared optics vanishes when the temperature goes to
zero, i.e., the Nernst heat theorem is satisfied. The same is proved
in \cite{PB43} for the Casimir entropy calculated with the
impedance of the anomalous skin effect. We stress that
at large separations the impedance
approach leads to magnitudes of the thermal Casimir force in
qualitative agreement with the case of ideal metals.

A critical problem of the impedance approach is the choice of the
functional dependence of the impedance function on the frequency.
In \cite{34} it was argued that the impe\-dan\-ce function valid in
the region around the characteristic frequency should be extrapolated to
lower frequencies, all the way to the zero Matsubara frequency.
In \cite{PB44} quantitative arguments were adduced in favor of
the statement that the use of different impedance functions within
different frequency regions in accordance with their applicability
conditions would be thermodynamically inconsistent. In
the first part of this paper we
apply the thermodynamic test to the impedance function of \cite{BiPRL}
which provides a smooth analytic interpolation between the impedances
of the normal skin effect and infrared optics. The substitution of the
interpolated impedance \cite{BiPRL} in the Lifshitz formula presents an
explicit example of the situation where different impedances are used
at Matsubara frequencies belonging to different frequency regions.
We present a rigorous analytic proof that the entropy of a fluctuating
field in this situation goes to a positive value when the temperature
vanishes. In other words, the Lifshitz formula combined with the
interpolated impedance is thermodynamically inconsistent.

The
problem of the thermal correction to the Casimir force was
further discussed
in \cite{Tor-Lam,BiCom,TLreply} where the impedance
of the normal skin
effect was used in the computations. In this approach
at a separation  of
1\,$\mu$m the thermal correction was found to be
about 30 times
larger than for ideal metal plates. This was
explained by the dominant
contribution of the transverse electric
evanescent waves of rather low
frequencies. The same impedance
function was applied \cite{BiPRL} to compute
the radiative heat
transfer through a vacuum gap between  two metal
surfaces at
different temperatures. This problem was considered
previously
in \cite{polder,loomis} (and recently in \cite{27})
using the
formalism of dielectric permittivity.
As proposed in \cite{BiPRL}, the
problems of the thermal Casimir
force and the radiative heat transfer
through a vacuum gap are
closely related and their simultaneous
investigation can be very
fruitful and elucidating.

In the second part of this paper we consider
both the thermal correction to the
Casimir force and the power of
radiative heat transfer using
different impedance functions and compare the results
with those obtained previously using the formalism of
dielectric permittivity and with experiment.
We try to address, from a purely phenomenological point of view,
the radiative heat transfer through a vacuum gap
in connection with the problem of
thermal Casimir force between real metals.
The aim of our work on this subject
is to come up with a phenomenological model for a real
metal at room temperature, that takes dissipation into account, and is
at the same time consistent with available experimental facts.
In view of the existing controversies among theoreticians about the
correct way to do this, it is our view that finding a reasonable
empirical model would be a valuable guide for further theoretical studies.
We perform computations in the framework of
two different formulations of   Lifshitz theory along the real
and the
imaginary frequency axis. This permits to specify the comparative
role of the
travelling and evanescent waves in the physical
phenomena under
consideration and to determine the application
range of different
approximations. It is shown that use of the
impedance of the normal skin effect
results in enormously large
thermal corrections to the Casimir force at
separations of about
200--300\,nm which are inconsistent with already
performed
experiments. As an alternative,  a phenomenological
generalized impedance of
infrared optics is constructed taking into account   relaxation
processes
specific for this frequency region which are not
connected with the
electron-phonon interactions. The thermal
correction to the Casimir force
computed with the generalized
impedance function is shown to be
qualitatively the same as is
known for ideal metals from thermal quantum
field theory in
Matsubara formulation. It is consistent with all
available
experimental data. The suggested impedance function is applied
to
calculate the power of radiative heat transfer between Au plates
at
different temperatures. At short separations between the plates
the obtained
power of   heat transfer per unit area is several
times less than the
one predicted previously in literature using
the dielectric function or
the impedance characteristic for the
region of the normal skin effect.
Thus we find that, depending on the chosen model of the metal,
one obtains largely different predictions for both
the thermal Casimir force and the power of
radiative heat transfer. This result underlines the
crucial role of new
experiments on the precision measurements of the
Casimir force and
the power of radiative heat transfer.

The paper is
organized as follows. In Sec.~2 we present   two equivalent
forms of Lifshitz formula, which involve evaluating the reflection
coefficients for imaginary and real frequencies, respectively.
In Sec.~3 we give the explicit expressions of the reflection
coefficients in
terms of  a frequency dependent
dielectric
permittivity and in terms of
Leontovich surface impedance.
In Sec.~4 we demonstrate that Lifshitz formula combined
with the impedance function of the infrared optics is thermodynamically
consistent. In Sec.~5 we prove that the substitution of the
interpolated impedance into the Lifshitz formula results in a violation
of the Nernst heat theorem.
Sec.~6 is devoted to the computation of
the thermal correction to the
Casimir force using different impedance
functions.
The results are compared with the case of plates made
of ideal
metal and with experiment. In Sec.~7 the general expression
for
the power per unit area of   radiative heat transfer
in terms of the
surface impedance is presented. Sec.~8 contains the
computation results
for   radiative heat transfer with different
impedances
and dielectric
permittivities including the new prediction to be
tested
experimentally. In Sec.~9 the reader will find  our conclusions
and
discussion.

\section{Lifshitz formula along the imaginary and real
frequency
axis}

In this section we consider two thick dissimilar plane parallel
plates
(semispaces) in thermal equilibrium at equal temperature
$T$,
separated by an empty gap of width $a$.
Let the $z$ axis be perpendicular to
the plates.
The Lifshitz formula
\cite{Lifshitz} represents the van der
Waals and Casimir free energy and
force per unit area (i.e., the pressure), acting
between the
plates marked by the indices (1) and (2),
in terms of the
reflection coefficients
$r_{\rm TM}^{(1,2)}(\omega,k_{\bot})$ and
$r_{\rm
TE}^{(1,2)}(\omega,k_{\bot})$
for two independent polarizations of
electromagnetic field
($\omega$ is the frequency and $k_{\bot}$ is the
magnitude
of the projection of the wave vector in the plane of the
plates).
The transverse magnetic polarization (TM) means that the
magnetic field
is perpendicular to the plane formed by
{\boldmath$k$}${}_{\bot}$ and
the $z$ axis, while for the
transverse
electric polarization (TE) the
electric field is perpendicular
to this plane. As was mentioned in the
Introduction,
in the literature there are many different
derivations of
the Lifshitz formula in the framework of
quantum statistical physics,
thermal quantum field theory in
Matsubara formulation and scattering
theory (see, e.g.,
\cite{4,5,6,7,8}). The final results of all
derivations
are represented in one of   two different forms,  as a
summation over
the Matsubara frequencies along the imaginary
frequency axis or,
alternatively, as an integral  over real
frequencies. We begin with the more
often used representation
in terms of imaginary frequencies, where the
Casimir free energy and pressure
are given
by
\begin{eqnarray}
&&
{\cal F}(a,T)=\frac{k_BT}{2\pi}\sum\limits_{l=0}^{\infty}
\left(1-\frac{1}{2}\delta_{l0}\right)\int_{0}^{\infty}
k_{\bot}dk_{\bot}
\label{PBeq3}
\\
&&\phantom{aaa}\times
\sum\limits_{\alpha={\rm
TE,TM}}\ln\left[1-
{r_{\alpha}^{(1)}(i\xi_l,k_{\bot})
r_{\alpha}^{(2)}(i\xi_l,k_{\bot})}e^{-2aq_l}\right],
\nonumber \\
&&
P(a,T)=-\frac{k_BT}{\pi}\sum\limits_{l=0}^{\infty}
\left(1-\frac{1}{2}\delta_{l0}\right)\int_{0}^{\infty}
k_{\bot}dk_{\bot}q_l
\label{equ-1}
\\
&&\phantom{aaa}\times
\sum\limits_{\alpha={\rm
TE,TM}}\left[
\frac{e^{2aq_l}}{r_{\alpha}^{(1)}(i\xi_l,k_{\bot})
r_{\alpha}^{(2)}(i\xi_l,k_{\bot})}-1\right]^{-1}.
\nonumber
\end{eqnarray}
\noindent
Here
$\xi_l=2\pi k_BTl/\hbar$ are the Matsubara frequencies,
$k_B$ is
Boltzmann constant, $l$ is a non-negative
integer
and
\begin{eqnarray}
&&
q(\omega,k_{\bot})=\sqrt{k_{\bot}^2-\frac{\omega^2}{c^2}},
\label{equ-2} \\
&&
q_l\equiv
q(i\xi_l,k_{\bot})=
\sqrt{k_{\bot}^2+\frac{\xi_l^2}{c^2}}.
\nonumber
\end{eqnarray}
\noindent
The
explicit form of the reflection coefficients
$r_{\alpha}^{(1,2)}$ is
discussed
in the next section.  Representation (\ref{equ-1})
is
convenient in numerical computations due to  fast
convergence of both the sum
and the integrals.

Using the Abel-Plana formula
\cite{8,43}
\begin{eqnarray}
&&
\sum\limits_{l=0}^{\infty}
\left(1-\frac{1}{2}\delta_{l0}\right)F(l)=
\int_{0}^{\infty}F(t)dt
\label{equ-3} \\
&&\phantom{aaaa}
+i\int_{0}^{\infty}dt
\frac{F(it)-F(-it)}{e^{2\pi
t}-1},
\nonumber
\end{eqnarray}
\noindent
where $F(z)$ is an
analytic function in the right half-plane,
Eqs.~(\ref{PBeq3}) and (\ref{equ-1}) can be
identically rearranged  in the form
\begin{eqnarray}
&&
{\cal F}(a,T)=E(a)+\Delta
{\cal F}(a,T), 
\nonumber \\
&&
P(a,T)=P_0(a)+\Delta
P(a,T).
\label{equ-4}
\end{eqnarray}
\noindent
Here $E(a)$ is given by
\begin{eqnarray}
&&
E(a)=\frac{\hbar}{4\pi^2}\int_{0}^{\infty}d\xi
\int_{0}^{\infty}k_{\bot}dk_{\bot}
\label{equ-4a} \\
&&\phantom{a}\times
\sum\limits_{\alpha={\rm
TE,TM}}\ln\left[1-
{r_{\alpha}^{(1)}(i\xi,k_{\bot})
r_{\alpha}^{(2)}(i\xi,k_{\bot})}e^{-2aq}\right]
\nonumber
\end{eqnarray}
\noindent
with
$q\equiv q(i\xi,k_{\bot})$ defined in Eqs.~(\ref{equ-2}).
The other
contribution on the right-hand side of the first equality
in Eq.~(\ref{equ-4})
can be represented in the form
\begin{equation}
\Delta
{\cal F}(a,T)=\frac{ik_BT}{2\pi}
\int_{0}^{\infty}dt
\frac{\Phi(i\xi_1t)-\Phi(-i\xi_1t)}{e^{2\pi t}-1},
\label{equ-4b}
\end{equation}
\noindent
where $\Phi(x)\equiv \Phi_{\rm TM}(x)+\Phi_{\rm TE}(x)$
and
\begin{eqnarray}
&&
\Phi_{\rm
TM,TE}(x)=\int_{0}^{\infty}k_{\bot}dk_{\bot}
\label{equ-4c} \\
&&\phantom{a}\times
\ln\left[1-
{r_{\rm TM,TE}^{(1)}(ix,k_{\bot})
r_{\rm TM,TE}^{(2)}(ix,k_{\bot})}
e^{2a\sqrt{k_{\bot}^2+\frac{x^2}{c^2}}}\right].
\nonumber
\end{eqnarray}
In  a similar way, the quantities $P_0(a)$ and
$\Delta P(a,T)$ in the second equality in Eq.~(\ref{equ-4})
are equal to
\begin{eqnarray}
&&
P_0(a)=-\frac{\hbar}{2\pi^2}\int_{0}^{\infty}d\xi
\int_{0}^{\infty}k_{\bot}dk_{\bot}q
\nonumber \\
&&\phantom{a}\times
\sum\limits_{\alpha={\rm
TE,TM}}\left[
\frac{e^{2aq}}{r_{\alpha}^{(1)}(i\xi,k_{\bot})
r_{\alpha}^{(2)}(i\xi,k_{\bot})}-1\right]^{-1}\, ,
\label{equ-5}\\
&&
\Delta
P(a,T)=-\frac{ik_BT}{\pi}
\int_{0}^{\infty}dt
\frac{F(i\xi_1t)-F(-i\xi_1t)}{e^{2\pi
t}-1},
\label{equ-6}
\end{eqnarray}
\noindent
where $F(x)\equiv F_{\rm TM}(x)+F_{\rm TE}(x)$
and
\begin{eqnarray}
&&
F_{\rm
TM,TE}(x)=\int_{0}^{\infty}k_{\bot}dk_{\bot}q
\sqrt{k_{\bot}^2+\frac{x^2}{c^2}}
\label{equ-7} \\
&&\phantom{a}\times
\left[
\frac{e^{2a\sqrt{k_{\bot}^2+
\frac{x^2}{c^2}}}}{r_{\rm
TM,TE}^{(1)}(ix,k_{\bot})
r_{\rm
TM,TE}^{(2)}(ix,k_{\bot})}-1\right]^{-1}.
\nonumber
\end{eqnarray}

Note that the quantities $E(a)$ and $P_0(a)$ in
Eqs.~(\ref{equ-4a}) and (\ref{equ-5}) are
often called in literature the Casimir energy and pressure at
zero
temperature, and $\Delta{\cal F}(a,T)$ and $\Delta P(a,T)$ in
Eqs.~(\ref{equ-4b}) and (\ref{equ-6}) are
referred to
as the thermal corrections to them. This
terminology is, however, correct
only for plate materials with
temperature independent properties. In
this case the Casimir free energy and
pressure depend on temperature only through the
Matsubara
frequencies and the thermal corrections defined
as ${\cal F}(a,T)-{\cal F}(a,0)$ and
$P(a,T)-P(a,0)$ coincide with $\Delta{\cal F}(a,T)$
and $\Delta P(a,T)$ in Eqs.~(\ref{equ-4b}) and (\ref{equ-6}).
If, however, the properties of a medium
(for instance, the dielectric permittivity)
depend on the
temperature, then the thermal corrections
${\cal F}(a,T)-{\cal F}(a,0)$ and $P(a,T)-P(a,0)$
do not coincide with $\Delta{\cal F}(a,T)$ and $\Delta P(a,T)$.
Even in this case
Eq.~(\ref{equ-4}) can be used to compute
the total Casimir free energy and pressure. In so doing, the
quantities $E(a)$ and
$P_0(a)$ in Eqs.~(\ref{equ-4a}) and (\ref{equ-5}) can be interpreted as
the contributions to the total Casimir free energy and pressure due to
zero-point oscillations [they
may depend on the temperature
as a parameter, i.e., in fact $E(a)=E(a,T)$ and
$P_0(a)=P_0(a,T)$] and
the quantities $\Delta{\cal F}(a,T)$ and $\Delta P(a,T)$ in
Eqs.~(\ref{equ-4b}) and (\ref{equ-6}) as
the contributions from thermal photons.

It is useful to express the Casimir free energy and
pressure as the integrals over
{\it real} frequencies
$\omega$:
\begin{eqnarray}
&&
{\cal F}(a,T)=\frac{\hbar}{4 \pi^2}\int_0^{\infty}d
\omega
\int_0^{\infty} d\,k_{\perp}\,k_{\perp} \, \coth\left(\frac{\hbar
\omega}{2 k_BT}\right)
\label{PBeq6} \\
&&\phantom{a}\times
{\rm Im}
\sum_{\alpha={\rm
TE,TM}}\ln\left[1-{r^{(1)}_{\rm\alpha}(\omega,k_{\bot})
r^{(2)}_{\rm\alpha}(\omega,k_{\bot})}
{e^{2\,i\,k_z\,a}}\right]\;,
\nonumber \\
&&
P(a,T)=-\frac{\hbar}{2 \pi^2}\int_0^{\infty}d
\omega
\int_0^{\infty} d\,k_{\perp}\,k_{\perp} \, \coth\left(\frac{\hbar
\omega}{2 k_B
T}\right)
\label{bi-1} \\
&&\phantom{a}\times
{\rm Re}
\left\{k_z\,\sum_{\alpha={\rm
TE,TM}}\left[1-\frac{e^{-2\,i\,k_z\,a}}{r^{(1)}_{\rm
\alpha}(\omega,k_{\bot})
r^{(2)}_{\rm
\alpha}(\omega,k_{\bot})}\right]^{-1}\right\}\;,
\nonumber
\end{eqnarray}
\noindent
where
\be
k_z(\omega,k_{\perp})\equiv\sqrt{\omega^2/c^2-k_{\perp}^2}
=iq(\omega,k_{\bot})\;.
\label{bi-2}
\ee
\noindent
We
note that real values of $k_z$ correspond to   propagating
waves
(PW),
while imaginary values of $k_z$ describe   evanescent waves
(EW).
Upon
using the identity
\be
\coth(x/2)=1+\frac{2}{\exp
(x)-1}\;,
\label{bi-3}
\ee
\noindent
we see that the  formula for the Casimir pressure can
be expressed
as the sum of two terms, like in
Eq.~(\ref{equ-4})
\begin{equation} P(a,T)=P_0(a,T)+\Delta P
(a,T)\;,
\label{bi-4}
\end{equation}
\noindent
where
\begin{eqnarray}
&&
P_0(a,T)=-\frac{\hbar}{2\pi^2}\int_0^{\infty}d
\omega
\int_0^{\infty} d\,k_{\perp} k_{\perp}\,
\label{bi-5}
\\
&&\phantom{a}\times
{\rm Re}
\left\{k_z
\sum_{\alpha={\rm
TE,TM}}\left[1-\frac{e^{-2\,i\,k_z\,a}}{r^{(1)}_{\rm
\alpha}(\omega,k_{\bot})
r^{(2)}_{\rm
\alpha}(\omega,k_{\bot})}\right]^{-1}\right\}\;,
\nonumber
\end{eqnarray}
and
\begin{eqnarray}
&&
\Delta
P(a,T)=-\frac{\hbar}{ \pi^2}\int_0^{\infty}d \omega
\int_0^{\infty}
d\,k_{\perp} k_{\perp}\,
\frac{1}{\exp\left(\frac{\hbar \omega}{k_B
T}\right)-1}
\nonumber \\
&&\times
{\rm
Re}
\left\{k_z\!\!\!\!\!\sum_{\alpha={\rm
TE,TM}}\!\left[1-\frac{e^{-2\,i\,k_z\,a}}{r^{(1)}_{\rm
\alpha}(\omega,k_{\bot})
r^{(2)}_{\rm
\alpha}(\omega,k_{\bot})}\right]^{-1}\!\right\}\!.
\label{bi-6}
\end{eqnarray}
\noindent
Similar representations can be easily obtained for $E(a,T)$
and $\Delta{\cal F}(a,T)$.
As was already noted above, $P_0(a,T)$ physically
represents the
contribution to the pressure from zero-point
fluctuations of the electromagnetic
field, and in general it
depends on the temperature, because the
permittivities
or surface impedances of the
plates are tempe\-ra\-tu\-re-dependent.
As
for $\Delta P(a,T)$, it represents the contribution from
thermally
excited electromagnetic fields, and it vanishes for $T=0$. For
this
reason, in what follows we shall conventionally  refer to
$\Delta P(a,T)$
as to the thermal correction to the Casimir
pressure. We further
consider the decomposition \be\Delta
P(a,T)=\Delta P_{\rm PW}(a,T)+\Delta
P_{\rm EW}(a,T)\;,
\label{bi-7} \ee \noindent where $\Delta P_{\rm
PW}(a,T)$ and
$\Delta P_{\rm EW}(a,T)$ represent the contributions from  PW
and
EW, respectively. It is easily seen from Eq. (\ref{bi-6}) that
$\Delta
P_{\rm PW}(a,T)$ can be written as:
\begin{eqnarray}
&&
\Delta P_{\rm
PW}(a,T)=-\frac{\hbar}{ \pi^2}\,\int_0^{\infty}d
\omega
\frac{1}{\exp\left(\frac{\hbar \omega}{k_B
T}\right)-1}\int_0^{\omega/c} d\,k_z \,k_z^2
\,
\nonumber
\\
&&\phantom{a}\times
\sum_{\alpha={\rm
TE,TM}}{\rm
Re}\left[1-\frac{e^{-2\,i\,k_z\,a}}{r^{(1)}_{\rm
\alpha}(\omega,k_{\bot})
r^{(2)}_{\rm
\alpha}(\omega,k_{\bot})}\right]^{-1}.
\label{bi-8}
\end{eqnarray}
\noindent
For EW it holds $k_{\bot}>\omega/c$ and
Im$(k_z)=q$.
As a result
$\Delta P_{\rm EW}(a,T)$ takes
the
form
\begin{eqnarray}
&&
\Delta P_{\rm EW}(a,T)=\frac{\hbar}{
\pi^2}\,\int_0^{\infty}d
\omega
\frac{1}{\exp\left(\frac{\hbar \omega}{k_B
T}\right)-1}
\int_0^{\infty} d\,q
\,q^2
\nonumber
\\
&&\phantom{a}\times
\sum_{\alpha={\rm
TE,TM}}{\rm
Im}\,\left[1-\frac{e^{2\,q\,a}}{r^{(1)}_{\rm
\alpha}(\omega,k_{\bot})
r^{(2)}_{\rm
\alpha}(\omega,k_{\bot})}\right]^{-1}.
\label{bi-9}
\end{eqnarray}
\noindent
It should be noted that $\Delta
P_{\rm EW}(a,T)$ vanishes in the case
of ideal metals, because for
$r_{\alpha}^{(1)}r_{\alpha}^{(2)}=1$
the quantity between square brackets in the above equation is
real. In fact, more
generally $\Delta P_{\rm EW}(a,T)$ vanishes whenever
the product of the
reflection coefficients $r_{\alpha}^{(i)}$ is real
and less
or equal to
unity, because then the quantity between the square
brackets in Eq.
(\ref{bi-9}) is real and has no zeroes. This is the
case for
example for TE EW, if the metal plates are described by the
plasma
model (see below).

\section{Reflection coefficients in terms of the
dielectric
permittivity and the surface impedance}

In   Lifshitz
theory the material media are described by
  dielectric permittivities  that
depend only on frequency
\cite{Lifshitz,DLP,LP-SPII}. The description
of the dielectric properties
of a medium by $\varepsilon(\omega)$ takes
full account
of   temporal dispersion but neglects
possible
contributions to the van der Waals and Casimir force from
  spatial dispersion.
In the formalism of the imaginary
frequency axis [see Eqs.~(\ref{PBeq3}) and (\ref{equ-1})] the reflection
coefficients are expressed in terms of the
dielectric
permittivity as follows:
\begin{eqnarray}
&&
r_{\rm
TM}^{(n)}(i\xi_l,k_{\bot})=
\frac{\varepsilon^{(n)}(i\xi_l)q_l-k_l^{(n)}}{\varepsilon^{(n)}(i\xi_l)q_l+
k_l^{(n)}},
\nonumber \\
&&
r_{\rm
TE}^{(n)}(i\xi_l,k_{\bot})=
\frac{k_l^{(n)}-q_l}{k_l^{(n)}+q_l},
\label{eq3-1}
\end{eqnarray}
\noindent
where
$n=1,\,2$ for the first and the second plates, respectively,
and
\begin{eqnarray}
&&
k^{(n)}(\omega,k_{\bot})=\sqrt{k_{\bot}^2-\varepsilon^{(n)}(\omega)
\frac{\omega^2}{c^2}},
\label{eq3-2}
\\
&&
k_l^{(n)}\equiv
k^{(n)}(i\xi_l,k_{\bot})=
\sqrt{k_{\bot}^2+\varepsilon^{(n)}(i\xi_l)
\frac{\xi_l^2}{c^2}}.
\nonumber
\end{eqnarray}

In
the formalism of real frequency axis used in Eqs.~(\ref{PBeq6}) and
(\ref{bi-1}) the
reflection coefficients are just (\ref{eq3-1}) but calculated
at real
frequencies
\begin{eqnarray}
&&
r_{\rm
TM}^{(n)}(\omega,k_{\bot})=
\frac{\varepsilon^{(n)}(\omega)q(\omega,k_{\bot})-
k^{(n)}(\omega,k_{\bot})}{\varepsilon^{(n)}(\omega)q(\omega,k_{\bot})+
k^{(n)}(\omega,k_{\bot})},
\nonumber
\\
&&
r_{\rm
TE}^{(n)}(\omega,k_{\bot})=
\frac{k^{(n)}(\omega,k_{\bot})-
q(\omega,k_{\bot})}{k^{(n)}(\omega,k_{\bot})+
q(\omega,k_{\bot})}\;.
\label{eq3-3}
\end{eqnarray}

The
reflection properties of electromagnetic waves on metal
surfaces are
often described in terms of the Leontovich surface
impedance
\cite{LLP-ED}. For isotropic metal surfaces the
surface impedance
relates the tangential components of electric
field and magnetic induction in the same way as in a plane wave
propagating in the interior of a metal perpendicular to its surface
\cite{LLP-ED}
\begin{equation}
{\mbox{\boldmath$E$}}_t=Z(\omega)\left[
{\mbox{\boldmath$B$}}_t\times{\mbox{\boldmath$n$}}\right],
\label{eq3-4}
\end{equation}
\noindent
Here {\boldmath$n$} is the unit vector normal to the
surface and
directed inside the medium. For an ideal metal $Z=0$.
 The boundary condition  (\ref{eq3-4}) is valid when
$|Z|\ll 1$. For good conductors this inequality is satisfied
within a wide frequency region.
Equation (\ref{eq3-4}) permits to
determine the electromagnetic field
outside the metal, without considering
the propagation of electromagnetic waves in the metal interior.
It is close in spirit to the original Casimir approach for
ideal metals and to so-called ``nonlocal'' boundary
condition \cite{EmigB} implied by a dielectric
permittivity depending only on the frequency. The latter
condition is in fact equivalent to the standard continuity
boundary conditions in classical electrodynamics but 
does not require
to consider field propagation inside a metal. Both the
standard continuity conditions and the ``nonlocal''
boundary condition are based on the spatially local relation
{\boldmath$D$}$(\mbox{\boldmath{$x$}},\omega)=
\varepsilon(\omega)\mbox{\boldmath$E$}(\mbox{\boldmath{$x$}},\omega)$
which assumes  space homogeneity. Therefore these boundary
conditions do not take into
account the effects of spatial dispersion. As a consequence,
the standard Lifshitz formula is applicable only in the
absence of spatial dispersion.  An advantage of the
surface impedance, as compared with
dielectric permittivity, is
that it permits \cite{Kats} to apply the Lifshitz formula
in the region of the anomalous skin effect, where the spatial
homogeneity is violated and the effects of spatial dispersion
should be taken into consideration. In \cite{kinetic} the
applicability of the condition (\ref{eq3-4}) in the region of
the anomalous skin effect is demonstrated from the solution of
kinetic equations. In this frequency region, a metal cannot be
characterized by a dielectric permittivity depending only
on frequency and, thus, the standard Lifshitz formula is not
applicable \cite{PRBcom}
(the Leontovich impedance, however, cannot be used at
short separations between the plates where the condition
$|Z|\ll 1$ is violated, see Introduction).
In the frequency regions where both quantities
$\varepsilon(\omega)$ and $Z(\omega)$ are well defined it holds
\begin{equation}
Z(\omega)=1/\sqrt{\varepsilon(\omega)}.
\label{eq3-4a}
\end{equation}

In optics of metals the reflection coefficients are usually
expressed in terms of
$Z(\omega)$ rather than $\varepsilon(\omega)$ \cite{LLP-ED}.
In \cite{Kats} the Leontovich surface impedance was used to
express the
reflection coefficients in the Lifshitz formula at
zero temperature. The
thermal Casimir force was presented in terms
of the surface impedance in
\cite{PB41}.
The derivation of the Lifshitz formula starting from impedance
boundary condition (\ref{eq3-4}) is contained in \cite{34}.
Importantly, as was
shown in \cite{34},
the use of the Leontovich impedance leads to
different results for
the thermal Casimir force between real metals, than
are obtained
by the use of the Drude dielectric function in
\cite{BS_00}.
The thermal
Casimir force computed within the impedance approach
was
demonstrated to be in qualitative agreement with the case of
ideal
metals and in accordance with the fundamentals of thermodynamics
and with
experiment. (The controversies between different
approaches to the
thermal Casimir force are  discussed in detail
in
\cite{BrPRE05,statusJPA06,wePRE06,BrJPA06}).

Using the Leontovich surface impedance instead of
the dielectric
permittivity, the reflection coefficients in
Eqs.~(\ref{PBeq3}) and (\ref{equ-1}) (in
the
formalism of the imaginary frequency axis) are given
by
\begin{eqnarray}
&&
r_{\rm
TM}^{(n)}(i\xi_l,k_{\bot})=
\frac{cq_l-Z^{(n)}(i\xi_l)\xi_l}{cq_l+Z^{(n)}(i\xi_l)\xi_l},
\nonumber \\
&&
r_{\rm
TE}^{(n)}(i\xi_l,k_{\bot})=
\frac{\xi_l-cq_lZ^{(n)}(i\xi_l)}{\xi_l+cq_lZ^{(n)}(i\xi_l)}.
\label{eq3-5}
\end{eqnarray}

In
the formalism of real frequency axis, the
reflection coefficients
expressed in terms of the Leontovich
impedance
are
\begin{eqnarray}
&&
r_{\rm
TM}^{(n)}(\omega,k_{\bot})=
\frac{ck_z(\omega,k_{\bot})-Z^{(n)}(\omega)\omega}{ck_z(\omega,k_{\bot})+
Z^{(n)}(\omega)\omega},
\label{eq3-6}
\\
&&
r_{\rm
TE}^{(n)}(\omega,k_{\bot})=
\frac{\omega-ck_{z}(\omega,k_{\bot})Z^{(n)}(\omega)}{\omega+
ck_{z}(\omega,k_{\bot})Z^{(n)}(\omega)},
\nonumber
\end{eqnarray}
\noindent
where
$k_z$ is defined in Eq.~(\ref{bi-2}).
An important problem arising in
the case of real metals is the
adequate choice of the functions
$\varepsilon(\omega)$ and
$Z(\omega)$.

The calculation of the surface impedance over the whole frequency
axis is based on kinetic theory \cite{kinetic}. Here we are interested
in two frequency regions. One of them is the infrared optics
defined by the inequalities
\begin{equation}
\frac{v_F}{\omega}\ll\delta_i\ll l(T),
\qquad
\omega\ll\omega_p,
\label{PBeq9}
\end{equation}
\noindent
where $v_F$ is the Fermi velocity, $l(T)$ is the mean free path of
a conduction electron, $\delta_i=c/\omega_p$ is the skin depth
and $\omega_p=2\pi c/\lambda_p$ is the plasma frequency.
The other frequency region of our interest is the region of the normal
skin effect characterized  by the inequalities
\begin{equation}
l(T)\ll\delta_N(\omega,T),
\qquad
l(T)\ll\frac{v_F}{\omega},
\label{PBeq10}
\end{equation}
\noindent
where $\delta_N(\omega,T)=c/\sqrt{2\pi\sigma_0(T)\omega}$ and
$\sigma_0(T)$ is the static electric conductivity.

In the frequency regions (\ref{PBeq9}) and (\ref{PBeq10}), both the
dielectric permittivity and the impedance have definite physical
meanings and are connected by Eq.~(\ref{eq3-4a}). In the region of
the infrared optics we have
\begin{equation}
\varepsilon_i(\omega)=1-\frac{\omega_p^2}{\omega^2},
\qquad
Z_i(\omega)=-i\frac{\omega}{\sqrt{\omega_p^2-\omega^2}}.
\label{PBeq11}
\end{equation}
\noindent
For the normal skin effect (\ref{PBeq10}) we have
\begin{eqnarray}
&&
\varepsilon_N(\omega)\equiv\varepsilon_N(\omega,T)=
i\frac{4\pi\sigma_0(T)}{\omega},
\nonumber \\
&&
Z_N(\omega)\equiv Z_N(\omega,T)=
(1-i)\sqrt{\frac{\omega}{8\pi\sigma_0(T)}}.
\label{PBeq12}
\end{eqnarray}

Let us consider two plates at room temperature $T=300\,$K
at the same separation distance of a few hundred nanometers, as in
\cite{BiPRL}. In this case, the respective characteristic frequency
of the Casimir effect $\Omega_c=c/(2a)$ belongs to the region of
infrared optics (\ref{PBeq9}). (As an example, at $a=400\,$nm,
$\Omega_c=3.75\times 10^{14}\,$rad/s and for gold
$\omega_p\approx 1.37\times 10^{16}\,$rad/s and
$v_F\approx 1.78\times 10^6\,$m/s.) In the Lifshitz formulas
(\ref{PBeq3}), (\ref{equ-1}) and (\ref{PBeq6}), (\ref{bi-1})
the frequencies of order $\omega_p$
and higher practically do not contribute to the result. All
Matsubara frequencies $\xi_l$ in Eqs.~(\ref{PBeq3}) and (\ref{equ-1}),
which contribute
to the result essentially, belong to the region of
infrared optics (\ref{PBeq9}) with the exception of $\xi_0=0$, which
belongs to the region of the normal skin effect (\ref{PBeq10})
(recall that $\xi_1=2.47\times 10^{14}\,$rad/s).
Real frequencies in Eqs.~(\ref{PBeq6}) and (\ref{bi-1}),
contributing to the result,
also belong to the region of infrared optics (\ref{PBeq9}) and to
the normal skin effect (\ref{PBeq10}).

In \cite{BiPRL} the
correlation functions of a fluctuating electromagnetic field
were expressed in terms of the surface impedance. The developed
formalism was applied to derive the Lifshitz formulas (\ref{PBeq6})
and (\ref{bi-1})
along the real frequency axis and to describe the radiation heat
transfer between two semispaces at different temperatures.
We consider the impedance function defined by the Drude
dielectric function which was used in \cite{BiPRL}
\begin{equation}
Z_D(\omega)=\frac{1}{\sqrt{\varepsilon_D(\omega)}},
\qquad
\varepsilon_D(\omega)=1-\frac{\omega_p^2}{\omega[\omega+i\gamma(T)]}.
\label{PBeq13}
\end{equation}
\noindent
Here $\gamma(T)$ is the relaxation parameter connected with the
above used parameters by the equation \cite{PB49}
\begin{equation}
\omega_p^2=4\pi\gamma(T)\sigma_0(T)=
\frac{4\pi\sigma_0(T)}{\tau(T)},
\label{PBeq14}
\end{equation}
\noindent
where $\tau=1/\gamma$ is the relaxation time.
In the region of infrared optics we have $\gamma(T)\ll\omega$.
As a result, $Z_D$ and $\varepsilon_D$ in Eq.~(\ref{PBeq13}) coincide
with $Z_i$ and $\varepsilon_i$ in Eq.~(\ref{PBeq11}), respectively.
At small frequencies, on the contrary, one can neglect $\omega$,
as compared to $\gamma$, and from Eq.~(\ref{PBeq14})
$Z_D$ and $\varepsilon_D$ in Eq.~(\ref{PBeq13}) coincide
with $Z_n$ and $\varepsilon_n$ in Eq.~(\ref{PBeq12}) describing
the frequency region of the normal skin effect. Therefore, Eq.~(\ref{PBeq13})
provides an expression of the impedance which is valid in
both frequency regions of the normal skin effect and infrared optics,
and it represents a smooth analytic interpolation between the two regions.

Below we demonstrate that the impedance of infrared optics, $Z_i$,
extrapolated to all lower frequencies, including zero frequency, leads
to zero entropy of a fluctuating field at $T=0$ (Sec.~4).
At the same time, the entropy of a fluctuating field calculated using
the impedance of the Drude model in Eq.~(\ref{PBeq13}) approaches a nonzero
positive value when the temperature vanishes (Sec.~5), hence
violating thermodynamics.

\section{Thermodynamic test for the surface impedance of
infrared optics}

We consider the free energy of a fluctuating field given by
Eqs.~(\ref{PBeq3}) and (\ref{PBeq6}) with reflection coefficients
(\ref{eq3-5}), (\ref{eq3-6}) and the impedance function (\ref{PBeq11}).
Our aim is to find the asymptotic behavior of the free energy and
entropy of a fluctuating field at low temperatures at separation
distances between two similar plates of a few hundred nanometers, so that
the characteristic frequency $\Omega_c$ belongs to the region of
infrared optics. The perturbation expansions in powers of the small
parameter $\kappa\equiv 4\pi k_BaT/(\hbar c)$ can be conveniently
carried out by using the dimensionless variables
\begin{equation}
\zeta_l\equiv\frac{\xi_l}{\Omega_c}=\frac{2a\xi_l}{c}=\kappa l,
\quad
y=2aq_l.
\label{PBeq15}
\end{equation}
\noindent
In terms of these variables the free energy of the fluctuating
field, Eq.~(\ref{PBeq3}), takes the form
\begin{eqnarray}
&&
{\cal F}(a,T)=\frac{\hbar c\kappa}{32\pi^2a^3}\sum\limits_{l=0}^{\infty}
\left(1-\frac{1}{2}\delta_{0l}\right)
\int_{\zeta_l}^{\infty}ydy
\label{PBeq16} \\
&&\phantom{aa}
\times\sum\limits_{\alpha={\rm TM,TE}}\ln\left[1-r_{\rm \alpha}^2(i\zeta_l,y
)e^{-y}\right].
\nonumber
\end{eqnarray}
\noindent
Using the variables (\ref{PBeq15}), the reflection coefficients (\ref{eq3-5})
are
\begin{eqnarray}
&&
r_{\rm TM}(i\zeta_l,y)=
\frac{y-Z(i\zeta_l\Omega_c)\zeta_l}{y+Z(i\zeta_l\Omega_c)\zeta_l}\, ,
\nonumber \\
&&
r_{\rm TE}(i\zeta_l,y)=
\frac{\zeta_l-yZ(i\zeta_l\Omega_c)}{\zeta_l+yZ(i\zeta_l\Omega_c)}\, .
\label{PBeq17}
\end{eqnarray}
\noindent
Here, the impedance function of infrared optics (\ref{PBeq11}) is
given by
\begin{equation}
Z(i\zeta_l\Omega_c)\equiv Z_i(i\zeta_l\Omega_c)=
\frac{\rho\zeta_l}{\sqrt{1+\rho^2\zeta_l^2}},
\label{PBeq18}
\end{equation}
\noindent
where $\rho\equiv\lambda_p/(4\pi a)=\delta_i/(2a)$ is much less than
unity throughout the entire region of application of the impedance approach.

In terms of new variables the Casimir energy at $T=0$ in Eq.~(\ref{equ-4a})
is given by
\begin{equation}
E(a)=\frac{\hbar c}{32\pi^2a^3}\int_{0}^{\infty}d\zeta
\int_{\zeta}^{\infty}f(\zeta,y)dy
\label{PBeq21}
\end{equation}
\noindent
and the thermal correction to it (\ref{equ-4b}) by
\begin{equation}
\Delta{\cal F}(a,T)=\frac{i\hbar c\kappa}{32\pi^2a^3}\int_{0}^{\infty}dt
\frac{\Phi(i\kappa t)-\Phi(-i\kappa t)}{e^{2\pi t}-1}\, ,
\label{PBeq22}
\end{equation}
\noindent
where the following notations are used:
\begin{eqnarray}
&&
f(\zeta,y)=y\ln\left[1-r_{\rm TM}^2(i\zeta,y)e^{-y}\right]
\nonumber \\
&&\phantom{aaaa}
+
y\ln\left[1-r_{\rm TE}^2(i\zeta,y)e^{-y}\right],
\nonumber \\
&&\Phi(x)\equiv\int_{x}^{\infty}dy\,f(x,y).
\label{PBeq23}
\end{eqnarray}

Now, we expand the function $f$ defined in Eq.~(\ref{PBeq23}) in powers of
the small parameter $\rho$ using Eqs.~(\ref{PBeq17}) and (\ref{PBeq18}):
\begin{eqnarray}
&&
f(x,y)=2y\ln(1-e^{-y})+4\rho\frac{x^2+y^2}{e^y-1}
\nonumber \\
&&\phantom{f(x,y)}
-8\rho^2\frac{(x^4+y^4)e^y}{y(e^y-1)^2}+\mbox{O}(\rho^3).
\label{PBeq24}
\end{eqnarray}

Substituting Eq.~(\ref{PBeq24}) into the definition of $\Phi$ in
Eq.~(\ref{PBeq23}), we obtain
\begin{equation}
\Phi(x)=I_0(x)+\rho I_1(x)+\rho^2I_2(x)+\mbox{O}(\rho^3),
\label{PBeq25}
\end{equation}
\noindent
where
\begin{eqnarray}
&&
I_0(x)=2\int_{x}^{\infty}ydy\ln(1-e^{-y}),
\nonumber \\
&&
I_1(x)=4\int_{x}^{\infty}dy\frac{x^2+y^2}{e^y-1},
\label{PBeq26} \\
&&
I_2(x)=-8\int_{x}^{\infty}dy\frac{(x^4+y^4)e^y}{y(e^y-1)^2}.
\nonumber
\end{eqnarray}

The integrals $I_0(x)$ and $I_1(x)$ are easily calculated \cite{PB50}
and are given by:
\begin{eqnarray}
&&
I_0(x)=-2\left[\mbox{Li}_3\left(e^{-x}\right)+
x\mbox{Li}_2\left(e^{-x}\right)\right]
\nonumber \\
&&\phantom{aaa}
=-2\zeta(3)+\frac{1}{2}x^2-x^2\ln x+\frac{1}{3}x^3+\mbox{O}(x^4),
\label{PBeq27} \\
&&
I_1(x)=8\left[\mbox{Li}_3\left(e^{-x}\right)+
x\mbox{Li}_2\left(e^{-x}\right)-x^2\ln\left(1-e^{-x}\right)\right]
\nonumber \\
&&\phantom{aaa}
=8\left[\zeta(3)-\frac{1}{4}x^2-\frac{1}{2}x^2\ln x+
\frac{1}{3}x^3+\mbox{O}(x^4)\right],
\nonumber
\end{eqnarray}
\noindent
where $\mbox{Li}_n(z)$ is the polylogarithm function and $\zeta(x)$
is the Riemann zeta function. As to the integral $I_2(x)$ in
Eq.~(\ref{PBeq26}), in is easily seen that
\begin{equation}
I_2(x)=-48\zeta(3)+\mbox{O}(x^4)
\label{PBeq28}
\end{equation}
\noindent
and, thus, $I_2(x)$ does not contribute to
$\Phi(i\kappa t)-\Phi(-i\kappa t)$ in the perturbation orders under
consideration.

Substituting Eq.~(\ref{PBeq27}) into Eq.~(\ref{PBeq25}), we obtain
\begin{eqnarray}
&&
\Phi(i\kappa t)-\Phi(-i\kappa t)=\pi i (\kappa t)^2-\frac{2}{3}\,i(\kappa t)^3
\label{PBeq29} \\
 &&\phantom{aaaaaaaaa}
+4\rho\left[\pi i (\kappa t)^2-\frac{4}{3}\,i(\kappa t)^3\right]
+\mbox{O}\left[(\kappa t)^4\right].
\nonumber
\end{eqnarray}

Now, it is easy to calculate the free energy of the fluctuating field
at small $\kappa$ from Eqs.~(\ref{equ-4}), (\ref{PBeq21}), (\ref{PBeq22})
and (\ref{PBeq29})
\begin{eqnarray}
&&
{\cal F}_i(a,T)=E(a)-\frac{\pi^2\hbar c}{720 a^3}\left[
\frac{45\zeta(3)}{8\pi^6}\kappa^3-\frac{1}{16\pi^4}\kappa^4\right.
\nonumber \\
&&\phantom{aaaaa}
\left.+\frac{45\rho}{\pi^4}\left(\frac{\zeta(3)}{2\pi^2}\kappa^3-
\frac{1}{90}\kappa^4\right)\right].
\label{PBeq30}
\end{eqnarray}
\noindent
It is convenient to introduce the so-called effective temperature
$k_BT_{\rm eff}=\hbar\Omega_c$ and to use the effective penetration
depth of electromagnetic oscillations into a metal (\ref{PBeq9})
in the frequency region of infrared optics,
$\delta_i=c/\omega_p=\lambda_p/(2\pi)$. Then, Eq.~(\ref{PBeq30}) can be
rearranged in the form
\begin{eqnarray}
&&
{\cal F}(a,T)=E(a)-\frac{\pi^2\hbar c}{720 a^3}\left\{
\frac{45\zeta(3)}{\pi^3}\left(\frac{T}{T_{\rm eff}}\right)^3-
\left(\frac{T}{T_{\rm eff}}\right)^4\right.
\nonumber \\
&&\phantom{aaaaa}
\left.+\frac{\delta_i}{a}\left[
\frac{90\zeta(3)}{\pi^3}\left(\frac{T}{T_{\rm eff}}\right)^3-
4\left(\frac{T}{T_{\rm eff}}\right)^4\right]\right\}.
\label{PBeq31}
\end{eqnarray}
\noindent
In the above, we have restricted our consideration to the second
perturbation order in the small parameter $\rho$. In the same way
as in \cite{PB51} it can be shown that the higher perturbation
orders in $\rho$ contain only terms of order $\mbox{O}(\kappa^n)$
with $n\geq 5$. It is notable that Eq.~(\ref{PBeq31}) coincides
\cite{36,37} with the free energy of the fluctuating field at low
temperatures obtained from the Lifshitz formula combined with the
dielectric permittivity of the plasma model $\varepsilon_i(\omega)$
in Eq.~(\ref{PBeq11}). Thus, the characterization of a metal by
means of the dielectric permittivity and Leontovich surface impedance
in the frequency region of infrared optics leads to the
same asymptotic behavior of the free energy at low temperatures.

The asymptotic behavior of the entropy of a fluctuating field
defined as
\begin{equation}
S(a,T)=-\frac{\partial{\cal F}(a,T)}{\partial T}
\label{PBeq32}
\end{equation}
\noindent
can be found by differentiating Eq.~(\ref{PBeq31})
\begin{eqnarray}
&&
S_i(a,T)=\frac{3k_B}{8\pi a^2}\,\left(\frac{T}{T_{\rm eff}}\right)^2
\left[\zeta(3)-\frac{4\pi^3}{135}\frac{T}{T_{\rm eff}}\right.
\nonumber \\
&&\phantom{aaaaa}
\left.+\frac{\delta_i}{a}\left(2\zeta(3)-
\frac{16\pi^3}{135}\frac{T}{T_{\rm eff}}\right)\right].
\label{PBeq33}
\end{eqnarray}
\noindent
As is seen in Eq.~(\ref{PBeq33}), entropy goes to zero when temperature
vanishes in accordance to the Nernst heat theorem. Thus, the Leontovich
impedance of the infrared optics withstands the thermodynamic test.
The Lifshitz formulas (\ref{PBeq3}) and (\ref{PBeq6}) combined with
the impedance of infrared optics are shown to be consistent with the
requirements of thermodynamics.

\section{Thermodynamic test for the surface impedance of
the Drude model}

In the previous section we have extrapolated the impedan\-ce function
of infrared optics to all lower frequencies, including zero
frequency. However, in the Lifshitz formula, Eq.~(\ref{PBeq3}), the zero
Matsubara frequency is situated outside the region of infrared optics.
In a similar manner, small real frequencies in the Lifshitz formula
(\ref{PBeq6}) satisfy inequalities (\ref{PBeq10}) and, thus, belong to
the region of the normal skin effect. Because of this, it seems
reasonable to use the impedance function (\ref{PBeq13}) which coincides
with the impedances of infrared optics and of normal skin effect
at high and low frequencies, respectively, and provides a smooth
analytic interpolation between the two frequency regions. In this
section we put the Lifshitz formulas, Eq.~(\ref{PBeq3})
and Eq.~(\ref{PBeq6}), combined with the
impedance of the Drude model (\ref{PBeq13}) to a thermodynamic test.

For $l\geq 1$ one can introduce the dimensionless variable
$x_l=\gamma(T)/\xi_l$ and rearrange the impedance of the Drude model
(\ref{PBeq13}) in the form
\begin{equation}
Z_D(i\xi_l)=Z_D(i\zeta_l\Omega_l)=
\frac{\rho\zeta_l}{\sqrt{\rho^2\zeta_l^2+\frac{1}{1+x_l}}}.
\label{PBeq34}
\end{equation}
\noindent
If the relaxation parameter $\gamma(T)$ is equal to zero, $x_l=0$ and
Eq.~(\ref{PBeq34}) coincides with Eq.~(\ref{PBeq18}) for the impedance
of the infrared optics. It is easily seen that for metals with
perfect crystal lattices $x_l\ll 1$ at sufficiently low $T$.
In fact, for $T=300\,$K for good metals it holds
$\gamma\sim 10^{13}-10^{14}\,$rad/s (as an example, for gold
$\gamma=5.32\times 10^{13}\,$rad/s), whereas $\xi_l=\xi_1l$ and
$\xi_1=2\pi k_BT/\hbar=2.46\times 10^{14}\,$rad/s leading to
$x_l=\gamma/\xi_l<0.22$. When $T$ decreases from $T=300\,$K to
approximately $T_D/4$, where $T_D$ is the Debye temperature
(for gold $T_D=165\,$K \cite{PB52}), $\gamma(T)$ decreases linearly,
$\gamma(T)\sim T$, i.e., following the same law as $\xi_l$.
At $T<T_D/4$ the relaxation parameter decreases according to the
Bloch-Gr\"{u}neisen law, $\gamma(T)\sim T^5$, due to electron-phonon
collisions \cite{PB49} and as $\gamma(T)\sim T^2$ at liquid helium
temperatures due to electron-electron scattering \cite{PB52}.
At $T=30\,$K and 10\,K it holds
$\gamma(T)/\xi_1(T)\approx 4.9\times 10^{-2}$ and $1.8\times 10^{-3}$,
respectively. The magnitude of parameter $x_l$ decreases further to zero
with $T\to 0$.

We represent the free energy of the fluctuating field in Eq.~(\ref{PBeq16})
in the form
\begin{equation}
{\cal F}_D(a,T)={\cal F}_D^{(l=0)}(a,T)+{\cal F}_D^{(l\geq 1)}(a,T),
\label{PBeq35}
\end{equation}
\noindent
where we separate the terms with zero and nonzero Matsubara
frequencies. Substituting the impedance function, Eq.~(\ref{PBeq13}), in
the reflection coefficients, Eq.~(\ref{PBeq17}), one obtains
\begin{equation}
r_{\rm TM}^2(0,y)=r_{\rm TE}^2(0,y)=1,
\label{PBeq36}
\end{equation}
\noindent
which leads to
\begin{equation}
{\cal F}_D^{(l=0)}(a,T)=\frac{k_BT}{8\pi a ^2}
\int_{0}^{\infty}y\,dy\ln\left(1-e^{-y}\right).
\label{PBeq37}
\end{equation}

The contribution from Matsubara frequencies with \hfill \\ $l\geq 1$,
${\cal F}_D^{(l\geq 1)}(a,T)$, is more cumbersome. We will
find its low-temperature asymptotic behavior perturbatively.
With this purpose we expand $Z_D$ in Eq.~(\ref{PBeq34}) in powers of
a small parameter $x_l$
\begin{eqnarray}
&&
Z_D(i\zeta_l\Omega_c)=\frac{\rho\zeta_l}{\sqrt{1+\rho^2\zeta_l^2}}
+\frac{(\rho\zeta_l)^{-2}x_l}{2[1+(\rho\zeta_l)^{-2}]^{3/2}}
+\mbox{O}(x_l^2)
\nonumber \\
&&\phantom{aaaaa}
=Z_i(i\zeta_l\Omega_c)+\tilde{x}_l+\mbox{O}(x_l^2),
\label{PBeq38}
\end{eqnarray}
\noindent
where the impedance of infrared optics, $Z_i$, is defined in
Eq.~(\ref{PBeq18}) and
\begin{equation}
\tilde{x}_l\equiv\frac{\gamma(T)}{2\omega_p}
(1+\rho^2\zeta_l^2)^{-3/2}\ll 1.
\label{PBeq39}
\end{equation}
\noindent
Now, we substitute Eq.~(\ref{PBeq38}) in Eq.~(\ref{PBeq16}) with omitted
zero-frequency contribution and expand ${\cal F}_D^{(l\geq 1)}(a,T)$
in powers of $\tilde{x}_l$ keeping only the first order term:
\begin{eqnarray}
&&
{\cal F}_D^{(l\geq 1)}(a,T)={\cal F}_i^{(l\geq 1)}(a,T)+
\frac{k_BT}{2\pi a^2}\sum\limits_{l=1}^{\infty}\zeta_l{\tilde{x}_l}
\label{PBeq40} \\
&&\phantom{a}\times
\int_{\zeta_l}^{\infty}y^2dye^{-y}
\left\{\frac{-y+\zeta_lZ_i(i\zeta_l\Omega_c)}{A_{\rm TM}}
+\frac{\zeta_l-yZ_i(i\zeta_l\Omega_c)}{A_{\rm TE}}\right\}
\nonumber \\
&&\phantom{aaaaaa}
+
\mbox{O}(\tilde{x}_l^2),
\nonumber \\
&&
\mbox{where}
\nonumber \\
&&
A_{\rm TM}=[y+
\zeta_lZ_i(i\zeta_l\Omega_c)]
\nonumber\\
&&\phantom{aaaa}\times
\left[\left(e^{-y}-1\right)\left(y^2+
\zeta_l^2Z_i^2(i\zeta_l\Omega_c)\right)
\right.
\nonumber \\
&&\phantom{aaaaaaaaaaaa}\left.
-2\left(e^{-y}+1\right)
\zeta_lyZ_i(i\zeta_l\Omega_c)\right]\, ,
\nonumber \\
&&
A_{\rm TE}=[\zeta_l+
yZ_i(i\zeta_l\Omega_c)]
\nonumber \\
&&\phantom{aaaa}\times
\left[\left(e^{-y}-1\right)\left(\zeta_l^2
+y^2Z_i^2(i\zeta_l\Omega_c)\right)
\right.
\nonumber \\
&&\phantom{aaaaaaaaaaaa}\left.
-2\left(e^{-y}+1\right)
\zeta_lyZ_i(i\zeta_l\Omega_c)\right].
\nonumber
\end{eqnarray}
\noindent
Here ${\cal F}_i^{(l\geq 1)}(a,T)$ is the free energy of the fluctuating
field, computed by using the impedance of infrared optics (\ref{PBeq18}),
with omitted contribution from the zero Matsubara frequency.
As a next step, we expand the integrand in Eq.~(\ref{PBeq40}) in powers of
a small impedance of infrared optics, $Z_i(i\zeta_l\Omega_c)$, and
retain only the zero order contribution [recall that
$Z_i(i\zeta_l\Omega_c)$ goes to zero when $T$ vanishes]:
\begin{eqnarray}
&&
{\cal F}_D^{(l\geq 1)}(a,T)={\cal F}_i^{(l\geq 1)}(a,T)
\nonumber \\
&&\phantom{a}+
\frac{k_BT}{2\pi a^2}\sum\limits_{l=1}^{\infty}{\tilde{x}_l}
\int_{\zeta_l}^{\infty}\frac{dy}{1-e^{y}}
\left(-\zeta_l+\frac{y^2}{\zeta_l}\right)
\nonumber \\
&&\phantom{a}
+\mbox{O}\left[Z_i(i\zeta_l\Omega_c)\tilde{x}_l,\tilde{x}_l^2\right].
\label{PBeq41}
\end{eqnarray}
\noindent
Bearing in mind the definition of $\tilde{x}_l$ in Eq.~(\ref{PBeq39}),
the sum in Eq.~(\ref{PBeq41}) takes the form
\begin{equation}
\Sigma=\frac{\gamma(T)}{2\omega_p}
\sum\limits_{l=1}^{\infty}(1+\rho^2\zeta_l^2)^{-3/2}
\int_{\zeta_l}^{\infty}\frac{dy}{1-e^{y}}
\left(-\zeta_l+\frac{y^2}{\zeta_l}\right).
\label{PBeq42}
\end{equation}
\noindent
Now we expand the integrand in Eq.~(\ref{PBeq42}) in powers of $e^{-y}$
and introduce the new variable $v=jy$, where
$j=1,\,2,\,3,\,\ldots\,$.
Thus, we obtain
\begin{equation}
\Sigma=\frac{\gamma(T)}{2\omega_p}
\sum\limits_{l=1}^{\infty}(1+\rho^2\zeta_l^2)^{-3/2}
\sum\limits_{j=1}^{\infty}\frac{1}{j}
\int_{j\zeta_l}^{\infty}dve^{-v}
\left(\zeta_l-\frac{v^2}{j^2\zeta_l}\right).
\label{PBeq43}
\end{equation}
\noindent
After the integration with respect to $v$, this leads to
\begin{equation}
\Sigma=-\frac{\gamma(T)}{\omega_p}
\sum\limits_{l=1}^{\infty}(1+\rho^2\zeta_l^2)^{-3/2}
\sum\limits_{j=1}^{\infty}\frac{1}{j^2}
\left(\frac{1}{j\zeta_l}+1\right)e^{-j\zeta_l}.
\label{PBeq44}
\end{equation}
\noindent
By expanding in powers of $\rho\zeta_l\equiv\rho\kappa l$ and
performing the summation in $l$ we obtain
\begin{eqnarray}
&&
\Sigma=\frac{\gamma(T)}{\omega_p}\left\{
\frac{1}{\kappa}\sum\limits_{j=1}^{\infty}\frac{1}{j^3}
\ln\left(1-e^{-j\kappa}\right)-
\sum\limits_{j=1}^{\infty}\frac{1}{j^2(e^{j\kappa}-1)}\right.
\nonumber \\
&&\phantom{a}
\left.+\frac{3}{2}\rho^2\kappa^2\left[
\frac{1}{\kappa}\sum\limits_{j=1}^{\infty}
\frac{e^{-j\kappa}}{j^3(1-e^{-j\kappa})^2}+
\sum\limits_{j=1}^{\infty}
\frac{e^{-j\kappa}(1+e^{-j\kappa})}{j^2(1-e^{-j\kappa})^3}\right]
\right.
\nonumber \\
&&\phantom{a}\left.
+
\mbox{O}(\rho^4\kappa^4)\right\}.
\label{PBeq45}
\end{eqnarray}
\noindent
In the asymptotic limit $\kappa\to 0$, Eq.~(\ref{PBeq45}) results in
\begin{equation}
\Sigma\approx\frac{\gamma(T)}{\omega_p}\,
\frac{\ln\kappa}{\kappa}\zeta(3)+
\frac{\gamma(T)}{\omega_p}\,\mbox{O}\left(\frac{1}{\kappa}\right).
\label{PBeq46}
\end{equation}
\noindent
The substitution of this expression into Eq.~(\ref{PBeq41}) leads to
\begin{eqnarray}
&&
{\cal F}_D^{(l\geq 1)}(a,T)={\cal F}_i^{(l\geq 1)}(a,T)+
\frac{k_BT\gamma(T)}{2\pi a^2\omega_p}\,
\frac{\ln\kappa}{\kappa}\zeta(3)
\nonumber \\
&&\phantom{a}
+\mbox{O}\left[Z_i(i\zeta_l\Omega_c)\tilde{x}_l,\tilde{x}_l^2\right]
+\frac{k_BT\gamma(T)}{2\pi a^2\omega_p}\,
\mbox{O}\left(\frac{1}{\kappa}\right).
\label{PBeq47}
\end{eqnarray}

Now we are in a position to find the asymptotic expression at small
$\kappa$ for the free energy computed using the impedance of the Drude
model. First we add the zero-frequency term
${\cal F}_D^{(l=0)}(a,T)$, defined in Eq.~(\ref{PBeq37}),
to both sides of Eq.~(\ref{PBeq47}). On the left-hand side of
this equation  the quantity ${\cal F}_D(a,T)$ is obtained. On the
right-hand side of Eq.~(\ref{PBeq47}) we add and subtract the
zero-frequency term of the free energy computed using the impedance
of the plasma model. This term is obtained from 
Eqs.~(\ref{PBeq16})--(\ref{PBeq18}):
\begin{eqnarray}
&&
{\cal F}_i^{(l=0)}(a,T)=\frac{k_BT}{16\pi a ^2}
\int_{0}^{\infty}y\,dy\left\{
\vphantom{\ln\left[\left(\frac{1-\rho y}{1+\rho y}\right)^2\right]}
\ln\left(1-e^{-y}\right)\right.
\nonumber \\
&&\phantom{aa}\left.
+
\ln\left[1-\left(\frac{1-\rho y}{1+\rho y}\right)^2\,e^{-y}\right]
\right\}.
\label{PBeq48}
\end{eqnarray}
\noindent
The term (\ref{PBeq48}), together with the first term on the
right-hand side of Eq.~(\ref{PBeq47}), gives us the free energy
${\cal F}_i(a,T)$ computed using the impedance of infrared optics.
As a result, from Eq.~(\ref{PBeq47}) it follows:
\begin{eqnarray}
&&
{\cal F}_D(a,T)={\cal F}_i(a,T)+\Delta{\cal F}_i^{(l=0)}(a,T)
\nonumber \\
&&\phantom{a}
+
\frac{k_BT\gamma(T)}{2\pi a^2\omega_p}\,
\frac{\ln\kappa}{\kappa}\zeta(3)
+\mbox{O}\left[Z_i(i\zeta_l\Omega_c)\tilde{x}_l,\tilde{x}_l^2\right]
\nonumber \\
&&\phantom{a}
+\frac{k_BT\gamma(T)}{2\pi a^2\omega_p}\,
\mbox{O}\left(\frac{1}{\kappa}\right),
\label{PBeq49}
\end{eqnarray}
\noindent
where
\begin{eqnarray}
&&
\Delta{\cal F}_i^{(l=0)}(a,T)={\cal F}_D^{(l=0)}(a,T)-
{\cal F}_i^{(l=0)}(a,T)
\label{PBeq50} \\
&&\phantom{aaaa}
=\frac{k_BT}{16\pi a^2}\,
\int_{0}^{\infty}y\,dy\left\{
\vphantom{\ln\left[\left(\frac{1-\rho y}{1+\rho y}\right)^2\right]}
\ln\left(1-e^{-y}\right)
\right.
\nonumber \\
&&\phantom{aaaaa}\left.
-
\ln\left[1-\left(\frac{1-\rho y}{1+\rho y}\right)^2\,e^{-y}\right]
\right\}
\nonumber \\
&&\phantom{aaaa}
=-\frac{k_BT}{16\pi a^2}\,
\left\{
\vphantom{\ln\left[\left(\frac{1-\rho y}{1+\rho y}\right)^2\right]}
\zeta(3)
\right.
\nonumber \\
&&\phantom{aaaaa}\left.
+\int_{0}^{\infty}y\,dy
\ln\left[1-\left(\frac{1-\rho y}{1+\rho y}\right)^2\,e^{-y}\right]
\right\}.
\nonumber
\end{eqnarray}
\noindent
Taking into account that $\kappa\sim T$ and at low temperatures
$\gamma(T)\sim T^2$, $Z_i(i\zeta_l\Omega_c)\sim T$,
$\tilde{x}_l\sim\gamma\sim T^2$, we arrive at the conclusion that
not only the last three terms on the right-hand side of
Eq.~(\ref{PBeq49}) vanish when temperature vanishes, but also their
derivatives with respect to temperature vanish.

Using Eq.~(\ref{PBeq32}), we can find  the asymptotic behavior of the
free energy and entropy of the fluctuating field at low temperatures
in the case that the  metal is described by the impedance of the Drude
model.
Keeping only the main terms of order $T$ and $T^2\ln T$
in Eq.~(\ref{PBeq49}) [recall that according to Eq.~(\ref{PBeq31})
${\cal F}_i(a,T)-E(a)\sim T^3$], we obtain
\begin{eqnarray}
&&
{\cal F}_D(a,T)=E(a)+\Delta{\cal F}_i^{(l=0)}(a,T)
\nonumber \\
&&
\phantom{aaaaaaa}+
\frac{k_BT\gamma(T)}{2\pi a^2\omega_p}\,
\frac{\ln\kappa}{\kappa}\zeta(3),
\label{PBeq51} \\
&&
S_D(a,T)=-\frac{\partial\Delta{\cal F}_i^{(l=0)}(a,T)}{\partial T}
\nonumber \\
&&\phantom{aaaaaaa}
-\frac{k_B\zeta(3)}{2\pi^2a^2}\,\frac{\gamma(T)}{\omega_p}\,
\frac{T_{\rm eff}}{T}\left(\ln\kappa+\frac{1}{2}\right)
\nonumber \\
&&\phantom{a}
=\frac{k_B}{16\pi a^2}\,
\left\{\zeta(3)+\int_{0}^{\infty}y\,dy
\ln\left[1-\left(\frac{1-\rho y}{1+\rho y}\right)^2\,e^{-y}\right]
\right\}
\nonumber \\
&&\phantom{aaaaaa}
-\frac{k_B\zeta(3)}{2\pi^2a^2}\,\frac{\gamma(T)}{\omega_p}\,
\frac{T_{\rm eff}}{T}\left[\ln\left(2\pi\frac{T}{T_{\rm eff}}\right)
+\frac{1}{2}\right].
\nonumber
\end{eqnarray}
\noindent
Expanding the integrand on the right-hand side of the second equality
in Eq.~(\ref{PBeq51})
in powers of $\rho$ and integrating with respect to $y$, we
arrive at
\begin{eqnarray}
&&
S_D(a,T)=\frac{k_B\zeta(3)}{2\pi a^2}\,\rho\left[1-6\rho+
\mbox{O}(\rho^2)\right]
\label{PBeq52} \\
&&\phantom{aaaaaa}
-\frac{k_B\zeta(3)}{2\pi^2a^2}\,\frac{\gamma(T)}{\omega_p}\,
\frac{T_{\rm eff}}{T}\left[\ln\left(2\pi\frac{T}{T_{\rm eff}}\right)
+\frac{1}{2}\right].
\nonumber
\end{eqnarray}
\noindent
To carry out the thermodynamic test of the impedance of the Drude model,
we consider $T$ approaching zero and find the following value of the
entropy:
\begin{eqnarray}
&&
S_D(a,0)=\frac{k_B}{16\pi a^2}\,
\left\{
\vphantom{\int_{0}^{\infty}y\,dy
\ln\left[\left(\frac{1-\rho y}{1+\rho y}\right)^2\right]}
\zeta(3)\right.
\nonumber \\
&& \phantom{aaaaa}\left.
+\int_{0}^{\infty}y\,dy
\ln\left[1-\left(\frac{1-\rho y}{1+\rho y}\right)^2\,e^{-y}\right]
\right\}
\nonumber \\
&&\phantom{aaaaaa}
=\frac{k_B\zeta(3)}{2\pi a^2}\,\rho\left[1-6\rho+
\mbox{O}(\rho^2)\right]>0.
\label{PBeq53}
\end{eqnarray}

As is seen from Eq.~(\ref{PBeq53}), the entropy of the fluctuating field
at zero temperature takes a nonzero positive value. This value
depends on the parameters of the system, i.e., on the separation
distance $a$, and on the plasma frequency $\omega_p$.
The latter participates through the definition of $\rho$
(recall that for metals without impurities described by
$\varepsilon_D$ the Casimir entropy at $T=0$ is negative \cite{36,37}).
What this means is that we have a violation of the third law of thermodynamics,
the Nernst heat theorem \cite{PB53}.

In the above calculations we have considered the entropy associated
with the fluctuating field. In other words, only the distance dependent
part of the total free energy was considered. However, inclusion of
the self-energies cannot invalidate our conclusion on the violation
of the Nernst heat theorem. The reason is that the entropy of a
fluctuating field in Eq.~(\ref{PBeq53}) depends on the separation
distance, whereas the entropies due to self-energies are
separation independent. As a consequence, the entropy of a fluctuating
field and the entropies of matter fields cannot cancel each other and
must satisfy the Nernst heat theorem separately.

Thus, the Leontovich impedance
of the Drude model is thermodynamically inconsistent and cannot be
used in combination with the Lifshitz formulas,
Eq.~(\ref{PBeq3}) and Eq.~(\ref{PBeq6}),
to calculate the thermal Casimir force. In the next section it is
also shown
that the thermal Casimir force computed using the impedance of
the Drude model is in disagreement with the experimental
data of \cite{Decca1}.
Thus, there are both theoretical and
experimental evidences against the use of this impedance function in the
theory of the thermal Casimir force.

\section{Alternative results for the thermal correction to
the
Casimir pressure computed with different impedance functions}

In
\cite{Tor-Lam,BiCom} the contribution from the transverse
electric
electromagnetic waves to the thermal correction \hfill \\
$\Delta P(a,T)$ was computed
using the real frequency axis
formalism and the impedance function of
the normal skin
effect $Z_N(\omega)$ defined in Eq.~(\ref{PBeq12}).
The
computation was performed for Au plates at room temperature
with   static conductivity
$\sigma_0=3\times 10^{17}\,\mbox{s}^{-1}$
and relaxation time $\tau=1.88\times
10^{-14}\,$s. As was noted in Sec.~3, the impedance
function (\ref{PBeq12}) can be obtained from
the dielectric
function in the region of the normal skin
effect, $\varepsilon_N(\omega)$, using Eq.~(\ref{eq3-4a})
where the static conductivity is connected with the
plasma frequency $\omega_p$ by Eq.~(\ref{PBeq14}) leading to
$\omega_p\approx 9.3\,$eV.
 In \cite{Tor-Lam,BiCom} the relatively
large
contribution to the thermal correction from the TE
waves  for plates
separation
$a=1\,\mu$m was obtained using the
impedance function (\ref{PBeq12}).
It is about 30 times larger than
the
corresponding correction for
ideal
metals. This was explained by the
increased role of the TE EW at
low
frequencies. According to \cite{TLreply} the predicted
increase of the
thermal correction at $a=1\,\mu$m is
in conflict with previous
theoretical work \cite{BS_00}
that predicts even several times  larger magnitudes
for
the thermal correction calculated using the dielectric
permittivity
$\varepsilon\sim\omega^{-1}$ at low frequencies.
Paper \cite{TLreply}
concludes also that the prediction of
\cite{Tor-Lam,BiCom} is
consistent with the experimental
data of \cite{Lam97} at $a=1\,\mu$m whereas
the
prediction of \cite{BS_00} is excluded by that
experiment.

Here we
repeat the numerical computations of \cite{Tor-Lam,BiCom}
for the
contribution of the TE mode  to the thermal correction to
the Casimir force,
using the impedance (\ref{PBeq12}). We also
compute the contribution
from the TM mode and study the role of EW
and PW in both contributions.
The computations are performed at
different separations using both
formalisms along the imaginary
and real frequency axis (presented in
Secs.~2 and 3) with practically
coinciding results. Although the thermal
correction predicted from
the impedance (\ref{PBeq12})  is consistent with
the
long-separation experiment \cite{Lam97} at 1$\,\mu$m,
it is shown to
be
excluded by the measurement of the Casimir force at
shorter
separations by means of the micromechanical torsional
oscillator
\cite{Decca1,Decca2,Decca3,D4}. Because of this, we also discuss several
other forms of impedance
function and find those consistent with
experiment.

For the purpose
of comparison with experiment (which is in fact
consistent with the
theoretical results for $P_0(a)$ at $T=0$
\cite{Decca1}), here we use
in
computations slightly different value of the plasma
frequency
$\omega_p=9.0\,$eV.
The thermal correction in the framework of the imaginary
frequency
axis formalism was computed by Eqs.~(\ref{equ-1}),
(\ref{equ-4})
and (\ref{equ-5}) for two similar Au plates at $T=300\,$K. The results
are
presented in Table~1 where column~1 contains the values of
the
separation distance between the plates and column~2 the values
of the thermal correction. Precisely the same values were
computed using Eq.~(\ref{bi-6}) of the real frequency axis formalism.
Note that although the magnitude of the thermal correction increases
with the decrease of separation, the relative thermal correction
becomes smaller at shorter separations. The same holds for metals
described by the dielectric permittivity of the plasma model
\cite{28,29}. In column~3 of Table~1 the ratios of
the thermal correction from column~2 to those for ideal
metal
(denoted by $\Delta P_{\rm TE}^{\rm IM}$)
are presented. Columns~4 and
5 contain the relative contributions
to the thermal correction from the
TE EW and PW, respectively,
computed by using Eqs.~(\ref{bi-8}) and
(\ref{bi-9}).
In columns~6 and 7 the relative contributions from the TM EW
and PW, respectively, are presented.
Here and below we perform all computations at $a\geq 200\,$nm
in order to remain well inside the application region of the
impedance approach. For example, at $a=200\,$nm the characteristic
frequency of the Casimir force is
$\omega_c=c/(2a)\approx 7.5\times 10^{14}\,$rad/s and
$Z_N(\omega_c)\approx 1.4\times 10^{-2}\ll 1$.
As is seen from columns~2 and 3 in
Table~1, the magnitudes of the total thermal correction computed with
the impedance function (\ref{PBeq12}) are rather large.
At separation
distance of 1$\,\mu$m the thermal correction for Au
plates is found to be
almost 16 times larger than for the plates
made of ideal metal. This
ratio quickly increases with the
decrease of separation. At a separation
of 200\,nm it is as
large as 5900. By the summation of the values
presented in
columns~4,\,5 from one hand and 6,\,7 from another hand, one
finds that the dominant contributions to $\Delta P$ are given by
the TE mode, whereas the relative contributions from the TM mode
are negligibly small. The largest value of the latter achieved at
$a=1\,\mu$m is equal
to 0.05. Comparing columns 4 and 5 from
one hand and 6 and 7 from
another, we can conclude that the
dominant contribution to $\Delta P_{\rm
TE}$
and $\Delta P_{\rm TM}$ is given by the EW.
If one considers only
the contribution from the TE mode discussed
in \cite{Tor-Lam}, one
obtains
$\Delta P_{\rm TE}/\Delta P_{\rm TE}^{\rm IM}=29.6$ and
$1.2\times
10^4$ at separations $a=1\,\mu$m and 0.2$\,\mu$m,
respectively.

Now we
compare the obtained magnitudes of the thermal correction
in column~2
of Table~1 with the experiment \cite{Decca1}.
For this purpose we need
to obtain the total magnitudes of the
Casimir pressure at temperature
$T=300\,$K.
The magnitudes of $P_0(a)$ can be most simply computed
using
Eqs.~(\ref{equ-5}) and (\ref{eq3-5}). However,   use of the
impedance
function (\ref{PBeq12}) leads to   incorrect values of
$P_0(a)$. As an
example, using (\ref{PBeq12}) one obtains
$|P_0|=716.5$ and 144.6\,mPa at
separations 200\,nm and 300\,nm,
respectively. At the same time, the
conventional magnitudes of
$P_0$ at these separations, obtained by
different authors
\cite{01,02,03} are 507.5 and 113.6\,mPa, respectively, in
drastic contradiction with the above values obtained by the use
of impedance (\ref{PBeq12}). The reason is that the dominant
contribution to
$P_0$ is given by the frequency region around
the characteristic
frequency $\omega_c$ which belongs not to the region of the
normal skin effect, where the
impedance function (\ref{PBeq12}) is appropriate, but to the
region of infrared optics.
On the contrary, the magnitude of the thermal correction
(\ref{bi-6})
with the impedance function (\ref{PBeq12}) is determined
by much lower frequencies where (\ref{PBeq12}) is applicable.

To compare
the thermal correction in column~2 of Table~1 with
experiment, we add it
to the conventional magnitudes of the
Casimir pressure at $T=0$
specified above. As a result, at
separations 200 and 300\,nm one obtains the
magnitudes of
the thermal Casimir pressure equal to 519.6 and
116.4\,mPa,
respectively. These should be compared with respective
measured
magnitudes of the Casimir pressure at the same separations equal
to 508.1
and 114.7\,mPa. The differences  between the above
theoretical and
experimental values are 11.5 and 1.7\,mPa.
They lie outside the boundary of
the half-width confidence
interval determined at 95\% confidence (at
separations 200 and
300\,nm the latter is equal to 8.6 and 1.6\,mPa,
respectively).
Thus, the impedance function (\ref{PBeq12})
is
not consistent with the measurement of the Casimir pressure at
short
separations by means of micromechanical torsional
oscillator
\cite{Decca1,Decca2,Decca3,D4}.

The impedance $Z_N$ is applicable only at low frequencies
specific
for the normal skin effect. As was discussed in Sec.~3, the
impedance
of the Drude model (\ref{PBeq13})
provides smooth interpolation between the regions of the normal
skin effect and infrared optics. At
all frequencies $\omega\ll\omega_p$ the impedance (\ref{PBeq13})
can be
represented in the form (\ref{PBeq12}) if  one replaces
$\sigma_0$ on
the right-hand side of (\ref{PBeq12}) by the ac
conductivity defined
as
\begin{equation}
\sigma(\omega)=\frac{\sigma_0}{1-i\tau\omega}.
\label{eq4-3a}
\end{equation}

It is easily seen that the impedance of the Drude model
(\ref{PBeq13})
leads to even larger magnitudes  for the thermal correction to
the
Casimir
pressure at short separations, than the impedance
(\ref{PBeq12}).
At the
separation $a=0.2\,\mu$m the magnitude of the thermal
correction
computed with the impedance (\ref{PBeq13}) is equal to
$|\Delta P|=14.7\,$mPa
and its ratio to the thermal correction for
ideal metals is 7200. At
$a=1\,\mu$m one obtains $|\Delta P|=0.0402\,$mPa
and $\Delta P/\Delta
P^{\rm ID}=20$. If  one considers only the
contribution from the TE mode,
one obtains
$\Delta P_{\rm TE}/\Delta P_{\rm TE}^{\rm ID}=37.4$ (close
to 36.5
obtained in \cite{BiCom} for a bit different value
of
$\omega_p=9.3\,$eV whereas in our computations here we use
$\omega_p=9.0\,$eV).
However, at $a\approx 200\,$nm this ratio
achieves the value
$\Delta
P_{\rm TE}/\Delta P_{\rm TE}^{\rm ID}\approx 14.3\times 10^3$.
As in the
case of impedance function (\ref{PBeq12}), the increase of
the thermal
correction is due to the contribution from the TE EW.

In the same way,
as before, it can be shown that the large thermal
correction predicted
by the impedance (\ref{PBeq13}) is excluded
experimentally. Thus, the
theoretical magnitudes
of the Casimir pressure obtained by
adding the
thermal correction to $P_0$ are equal to 522.15 and
116.99\,mPa at
separations 200 and 300\,nm, respectively.
By comparing this with experiment
we get the respective differences
14.06 and 2.32\,mPa which are far
outside the boundary of the 95\%
confidence interval presented
above.
[Notice that the use of the Drude dielectric function
from Eq.~(\ref{PBeq13})
with reflection coefficients
(\ref{eq3-1}) to compute the
thermal
Casimir force is inconsistent not only with experiment
\cite{Lam97} but with
experiments \cite{18,Decca1,Decca2,Decca3,D4} as well.]

Bearing in mind that in
the separation range from 200\,nm to 1\,$\mu$m
the characteristic
frequency $\Omega_c=c/(2a)$ belongs to the region
of infrared optics, it is
reasonable to calculate the thermal
correction to the Casimir pressure
using the impedance $Z_i$
defined in Eq.~(\ref{PBeq11}).
Note
that the dielectric permittivity $\varepsilon_i(\omega)$ with
the
reflection coefficients (\ref{eq3-1}) was first used to calculate
the
thermal Casimir pressure in \cite{28,29}.
It was shown that the dielectric
permittivity of the plasma model
leads to small thermal corrections to
the Casimir pressure at short
separations in qualitative agreement with
the case of ideal metals.
At large separations it leads to the same
result as for
ideal metals in accordance to the classical limit
\cite{Rez,Jaffe}.

The disadvantage of the impedance function (\ref{PBeq11}) is
that
it has zero real part at nonzero frequencies, i.e., does not take
relaxation processes into account. At the same time the
tabulated
optical data for the complex index of refraction  $n$ in the region
of infrared optics \cite{Palik} lead to a nonzero real part of the
impedance function which, however, is much less than the magnitude
of the imaginary part.
Unfortunately, the optical data at low frequencies are not
available. In addition, these data are burdened by large errors and
uncertainties. Because of this, any theoretical result for the
functional form of the impedance function is of much value.
Preserving only the first expansion orders for
both real and
imaginary parts,  a more accurate complex impedance
in the region of the
infrared optics can be approximately presented in the
form
\cite{LLP-ED}
\begin{equation}
{Z}_p(\omega)=C\omega^2-i\frac{\omega}{\sqrt{\omega_p^2-\omega^2}},
\label{eq4-5}
\end{equation}
\noindent
where
the constant $C=0.004\,\mbox{eV}^{-2}$ was chosen to provide
the best
mean fit to the optical data for the real part of the impedance
within
the frequency region from 0.125\,eV to 5\,eV.
The impedance (\ref{eq4-5}) disregards interband transitions
but the real part of it takes into account electron-electron
collisions which occur rather seldom in the region of infrared
optics.

In Fig.~1 the imaginary
(a) and real (b) parts of the impedance
(\ref{eq4-5}) are shown by the
solid lines as functions of the
frequency. In the same figure the
imaginary (a) and real (b) parts of
the impedance are plotted by dots using
the expression
$Z_p(\omega)=1/n(\omega)$ and the tabulated optical
data
for $n(\omega)$ \cite{Palik}. As is seen in Fig.~1b, in the
region
$\omega\leq 1.5\,$eV the analytic expression for the imaginary part
of the
impedance (\ref{eq4-5}) is in rather good agreement with data.
This
expression gives the major contribution to the impedance along
the
imaginary frequency
axis
\begin{equation}
{Z}_p(i\xi)=-C\xi^2+\frac{\xi}{\sqrt{\omega_p^2+\xi^2}}.
\label{eq4-6}
\end{equation}
\noindent
Then one
can conclude that the approximation (\ref{eq4-6}) is well
adapted for the
computation of the thermal correction to the
Casimir pressure.
\begin{figure}[t]
\vspace*{-1.5cm}
\hspace*{-3cm}
\resizebox{0.75\textwidth}{!}{%
 \includegraphics{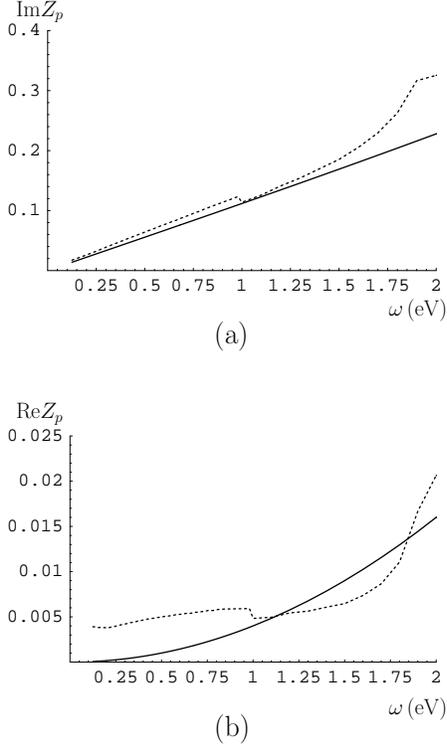}
}
\vspace*{-7.cm}
\caption{Imaginary
(a) and real (b) parts
of the impedance of infrared optics as functions
of frequency.
Solid lines are for the analytical expression (73),
dots
correspond to the tabulated optical
data.}
\label{fig:1}       
\end{figure}

The
computations were performed for Au plates by using both the
imaginary and
real frequency axis formalisms discussed in Sec.~2
with
practically
coinciding results for the total thermal correction and also
for separate
contributions from the TE and TM modes. When using
the real frequency
formalism, some care is necessary in the
numerical
integration of
Eqs.(\ref{bi-8}) and (\ref{bi-9}), because, due to the smallness of
the real part of $Z_i$,
there exist narrow resonances in the
spectrum of the thermal correction,
corresponding to the
eigenfrequencies of a dissipationless cavity with
impedance
${Z}_i(\omega)$. These considerations apply especially
to
the computation  $\Delta P_{\rm{TM, EW}}$, because its spectrum
has
resonances also at thermal frequencies. These results are
presented in
Table~2. In column~2 the magnitudes of the total
thermal correction to
the Casimir pressure at different
separations are listed. Column~3
contains the ratios of the total
thermal corrections for Au plates to those
for   ideal metals. In
column~4 the relative contributions from the TE
EW to the total
thermal correction are shown. Column~5 contains the
relative
contributions from the TE PW to the total thermal
correction.
Columns 6 and 7 contain the relative contributions to the total
thermal
correction from the TM EW and TM PW, respectively. As is
seen from
columns 2 and 3 in Table~2, the impedance of
infrared optics leads to much
smaller thermal corrections than the
impedance of the normal skin effect
in qualitative agreement with
the case of ideal metals. From columns
4--7 it follows that the TE
and TM modes lead to qualitatively similar
contributions. The role
of EW is also not so pronounced as it was for the
impedance of the
normal skin effect.

The total thermal Casimir
pressures obtained by the summation of the
thermal corrections in column~2 of
Table~2 and zero-temperature
pressures are consistent with all
available experimental data.
By way of example, at $a=200$ and 300\,nm (where
for the impedance
of the normal skin effect the theoretical results were
excluded at
95\% confidence) the differences between theory and
experiment
\cite{Decca1,Decca2} are now equal to --0.61 and
--1.1\,mPa,
respectively, i.e., well inside the 95\% confidence interval.
Note that the thermal Casimir force computed using the dielectric
permittivity (\ref{PBeq11}) is also consistent with data
\cite{Decca1}. The addition of a small imaginary part to
$\varepsilon_i(\omega)$, arising from real part of the
impedance in Eq.~(\ref{eq4-5}), leads to only minor changes in
the computation results which remain consistent with experiment.

Thus, the
impedance of infrared optics (\ref{eq4-6}) leads to
reasonable results
when applied to the thermal Casimir pressure.
In the following sections we
are going to apply the impedance
method to the computation of the
radiative heat transfer across a
vacuum gap between  two parallel surfaces
at different
temperatures. As is shown in Sec.~7 below, this process is
mostly
determined by the real part of the impedance function which
is
modelled not enough precisely in Eq.~(\ref{eq4-5}) (see
Fig.~1b).
Bearing in mind the computations of the radiative heat transfer,
we introduce
one more model impedance of  infrared optics by
the
equation
\begin{eqnarray}
&&
Z_t(\omega)=
\left\{
\begin{array}{ll}
B\sin\left(\frac{\pi\omega^2}{2\beta^2}\right),\quad
&
\omega\leq\beta,\\
B,&\beta\leq\omega\leq 0.125\,\mbox{eV},
\\
Y(\omega), & \omega\geq
0.125\,\mbox{eV}
\end{array}\right\}
\nonumber \\
&&\phantom{aaa}
-i\frac{\omega}{\sqrt{\omega_p^2-\omega^2}}.
\label{eq4-7}
\end{eqnarray}
\noindent

Here $Y(\omega)$ stands for the tabulated optical data
representing the
real part of the impedance \cite{Palik} which are
available at
$\omega\geq 0.125\,$eV. The value of the constant
$B=0.00389$ is chosen such as
to  have a   smooth transition
between the
optical data and their
extrapolation to lower frequencies. The
unknown parameter $\beta$ fixes the
value of frequency where the
behavior of Re$Z_t$, as given by the
optical data, is smoothly
connected with the asymptotic behavior at low frequencies.
The upper bound of $\beta$ is determined by the fact that the
optical data are available at $\omega\geq 0.125\,$eV.
In the frequency region of infrared optics the real
part must be much smaller than the imaginary.
The reason is that at these frequencies electrons are
almost free and electric current is pure imaginary.
The small real part of the impedance describes minor
distortions in the vibrational motion of electrons. Then
it is reasonable to impose the lower constraint
on $\beta$ as follows: $0.08\,\mbox{eV}\leq\beta$.
This constraint is very conservative because it allows
the real part of the impedance to become as large as one half of
the imaginary part (a larger real part of the impedance in the
region of infrared optics is evidently inadmissible).
The real part of the
impedance (\ref{eq4-7}) as a function of frequency is shown in
Fig.~2, where
line 1 corresponds to $\beta=0.125\,$eV and line 2
to $\beta=0.08\,$eV.
The region between   lines 1 and 2 is
allowed.
\begin{figure}[t]
\vspace*{-6cm}
\hspace*{-2.5cm}
\resizebox{0.75\textwidth}{!}{%
 \includegraphics{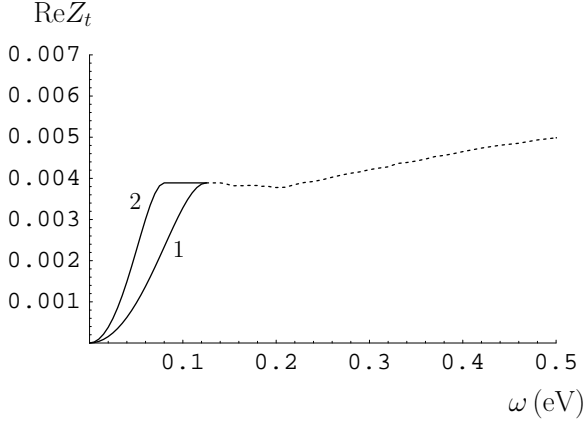}
}
\vspace*{-7cm}
\caption{Real
part
of the impedance of infrared optics given by the
tabulated optical
data (dots) versus frequency
with different extrapolations to low
frequencies
(the region between the solid lines 1 and 2).
See text for
further
discussion.}
\label{fig:2}
\end{figure}

\section{Heat transfer across an empty gap}

We
consider now the case of two semi-infinite plane parallel
metallic
plates at different temperatures $T_1>T_2$, separated by
an empty gap of
width $a$. We are interested in estimating the
heat transfer between the
plates. This problem was first studied
long ago by Rytov \cite{rytov},
and later on it was reconsidered
by Polder and Van Hove \cite{polder},
by Loomis and Maris
\cite{loomis} and by Volokitin and Persson
\cite{27}.
Recently, it was studied by one of us \cite{BiPRL}
in the framework
of the surface impedance.  The approach followed
by Polder and Van Hove
is closely related to Lifshitz theory of
the van der Waals interactions
between macroscopic bodies, in that
heat transfer is regarded as
occurring via fluctuating
electromagnetic fields radiated by the two plates,
whose sources
are the random thermal electric currents that are present
inside
the plates. The fluctuation-dissipation theorem is then used
to
determine the statistical properties of the currents, from which
the
correlation properties of the emitted electromagnetic fields
are
subsequently derived. In the derivation, Polder and Van Hove
made the following
assumptions, that are also at the basis of
Lifshitz theory: i) the
wavelengths of the electromagnetic fields
involved should be large compared
to atomic distances, so that the
fluctuating fields can be well
described by means of the classical
macroscopic Maxwell's equations; ii) the
electric currents at
distinct points inside the plates are uncorrelated;
iii) the media
are isotropic and nonmagnetic, and such that their
electromagnetic
properties can be described by means of a complex
dielectric
permittivity
$\varepsilon(\omega)$ that depends only on the frequency;
iv) the
system is stationary in time, and each plate is in local
thermal
equilibrium. The resulting expression for the power (per unit
area)
$S$ of heat transfer from plate one to plate two,
was found using the average value of the Poynting vector in the
gap between two plates
\cite{loomis}:
\begin{eqnarray}
&&
S=\frac{4
\,\hbar}{\pi^2}
\int_0^{\infty}\,d\omega \,\omega 
 \int_0^{\infty} dk_{\perp} \,k_{\perp}\, k_z^2
|e^{2 i k_z
a}|
\label{heattr} \\
&&\phantom{a}
\nonumber \\
&&\phantom{a}\times
\left(\frac{1}{\exp(\hbar\omega/k_B T_1)-1}
-\frac{1}{\exp(\hbar \omega/k_B T_2)-1}\right)
\nonumber \\
&&
\phantom{a}
\times\left[\frac{{\rm Re}(s^{(1)})\, 
{\rm Re}(s^{(2)})}{X_{\rm TE}} +
\frac{{\rm Re}(\bar{\varepsilon}^{(1)}\,s^{(1)})\,
{\rm
Re}(\bar{\varepsilon}^{(2)}\,s^{(2)})}{X_{\rm TM}}\right]\;,
\nonumber \\
&&
\mbox{where}
\nonumber \\
&&
X_{\rm TE}=|(k_z+s^{(1)}) (k_z+s^{(2)})
\nonumber \\
&&\phantom{aaa}
-(k_z-s^{(1)})
(k_z-s^{(2)})e^{2 i k_za}|^2\, ,
\nonumber \\
&&
X_{\rm TM}=|(\varepsilon^{(1)}\,k_z+s^{(1)})
(\varepsilon^{(2)}\,k_z+s^{(2)})
\nonumber \\
&&\phantom{aaa}
-(\varepsilon^{(1)}\,k_z-s^{(1)})
(\varepsilon^{(2)}\,k_z-s^{(2)})e^{2i k_z a}|^2\, ,
\nonumber
\end{eqnarray}
\noindent
and
\begin{equation}
s^{(n)}(\omega,k_{\bot})=
\sqrt{\varepsilon^{(n)}(\omega)\,\omega^2/c^2-k_{\perp}^2}
=ik^{(n)}(\omega,k_{\bot})
\label{eq5-1a}
\end{equation}
\noindent
(note
that
our variable $k_{\perp}$ is denoted by $q$ in \cite{27,loomis}).
We now
decompose $S$ as the sum of the contributions from PW and
EW:
\be
S=S_{\rm PW}\,+\,S_{\rm EW}\;.
\label{splitht}
\ee
\noindent
It is
not hard
to verify, starting from  Eq. (\ref{heattr}), that
$S_{\rm PW}$ and
$S_{\rm EW}$ can be expressed in terms of the
dielectric reflection
coefficients (\ref{eq3-3}) as \cite{27}:
\begin{eqnarray}
&&
S_{\rm
PW}=\frac{\hbar}{4 \pi^2} \int_0^{\infty}\,d\omega
\omega\,\int_0^{\omega/c} d
k_z\;k_z\,
\label{heatpw} \\
&&\phantom{a}
\times\left(\frac{1}{\exp(\hbar \omega/k_B T_1)-1}-
\frac{1}{\exp(\hbar \omega/k_B
T_2)-1}\right)
\nonumber \\
&&\phantom{a}
\times\sum_{\alpha={\rm
TE,TM}}\frac{\left[1-|r_{\alpha}^{(1)}(\omega,k_{\bot})|^2\right]
\left[1-|r_{\alpha}^{(2)}(\omega,k_{\bot})|^2\right]}
{|1-r_{\alpha}^{(1)}(\omega,k_{\bot})
r_{\alpha}^{(2)}(\omega,k_{\bot})\exp(2i\,
k_z\,a)|^2}\,,
\nonumber\\
&&
S_{\rm EW}=\frac{\hbar}{\pi^2}
\int_0^{\infty}\,
d\omega \,\omega\,\int_0^{\infty}
dq\;q\,
\label{heatew} \\
&&\phantom{a}
\times\left(\frac{1}{\exp(\hbar \omega/k_B
T_1)-1}-
\frac{1}{\exp(\hbar \omega/k_B T_2)-1}\right)
\nonumber
\\
&&\phantom{a}
\times\sum_{\alpha={\rm
TE,TM}}\frac{ {\rm Im}r_{\alpha}^{(1)}(\omega,k_{\bot})\,
{\rm
Im}r_{\alpha}^{(2)}(\omega,k_{\bot})
\,e^{-2\,q\,a}}
{|1-r_{\alpha}^{(1)}(\omega,k_{\bot})
r_{\alpha}^{(2)}(\omega,k_{\bot})\exp(-2\,q\,a)|^2}
\;.
\nonumber
\end{eqnarray}
\noindent
It
is to be noted that, for
$r_{\alpha}^{(2)}=0$, $S_{\rm EW}=0$, while
for
$r_{\alpha}^{(2)}=0$, and $T_2=0$ the expression for $S_{\rm
PW}$
reduces to the well known Kirchhoff's formula for the flux of
radiation
$\Phi$ from a surface with reflection
coefficients
$r_{\alpha}=r_{\alpha}^{(1)}$, at temperature $T=T_1$:
\begin{eqnarray}
&&
\Phi(T)=\frac{1}{4 \pi^2\, c^2}
\int_0^{\infty}\,d\omega
\,\frac{\hbar \omega^3}{\exp(\hbar \omega/k_B
T)-1}
\nonumber \\
&&\phantom{aaaa}\times
\int_0^{1}
dp\;p\,\sum_{\alpha={\rm
TE,TM}}(1-|r_{\alpha}|^2)\;,
\label{kirch}
\end{eqnarray}
\noindent
where $p=k_z\,
c/\omega$.

In \cite{BiPRL}
the power of heat transfer per unit area $S$ was
computed within a
general theory of electromagnetic fluctuations for metallic
surfaces, based
on the concept of surface impedance. This approach
is closer to that
originally followed by Rytov \cite{rytov}, in
that the starting point of
the theory is an expression for the
correlators of the electric and
magnetic fields. However,
according to the dictate of  impedance theory,
in \cite{BiPRL}
the correlators  are given only {\it outside} the metal,
while no
consideration is made of either the fields, or the
electric
currents in the {\it interior} of the metal. The resulting
expression for
the heat transfer that was found in
\cite{BiPRL} is:
\begin{eqnarray}
&&
S=\frac{4 \hbar c^2}{\pi^2} \int_0^{\infty} \frac{d
\omega}{\omega}
\,
{{\rm
Re}Z^{(1)}(\omega)\,
{\rm Re}Z^{(2)}(\omega) }
\nonumber \\
&&\phantom{aa}\times
\left(\frac{1}{\exp (\hbar \omega/ k_B
T_1)-1}-\frac{1}{\exp (\hbar
\omega/ k_B T_2)-1}\right)
\nonumber \\
&&\phantom{aa}
\times
\int_0^{\infty} d
k_{\perp}
\,k_{\perp}\,|k_{z}|^2 \,|e^{2 i
k_z\,a}|\left(\frac{1}{B_{TE}}+
\frac{1}{B_{TM}}\right)\, ,
\label{htim}
\end{eqnarray}
\noindent
where the
quantities $B_{TE/TM}$ are defined
as:
\begin{eqnarray}
&&
B_{TE}=|(1+k_z\,c \,Z^{(1)}/\omega)(1+ k_z\,c \,Z^{(2)}/\omega)
\label{eq5-7}
\\
&&\phantom{a}-(1-
k_z\,c \,Z^{(1)}/\omega)(1- k_z\,c \,Z^{(2)}/\omega)
\exp (2 i
k_z\,a)|^2\;,
\nonumber \\
&& B_{TM}=|(k_z\,c/\omega+
Z^{(1)})(k_z\,c/\omega+
Z^{(2)})
\label{eq5-8}
\\
&&\phantom{a}
-(k_z\,c/\omega-Z^{(1)})(k_z\,c/\omega-Z^{(2)}) \exp (2
i
k_z\,a)|^2\;.
\nonumber
\end{eqnarray}
\noindent
If we separate the PW and EW
contributions to
Eq.~(\ref{htim}), it is easy to verify
that
Eqs.~(\ref{splitht})--(\ref{heatew})
are  equivalent to Eqs.~(\ref{htim})--(\ref{eq5-8})
 in
the impedance theory of
heat transfer, provided that the reflection
coefficients are taken
to be those of impedance theory in
Eq.~(\ref{eq3-6}).

\section{Numerical results for heat transfer and emittivity}

\begin{figure}[t]
\vspace*{-3.7cm}
\hspace*{-3cm}
\resizebox{0.75\textwidth}{!}{%
 \includegraphics{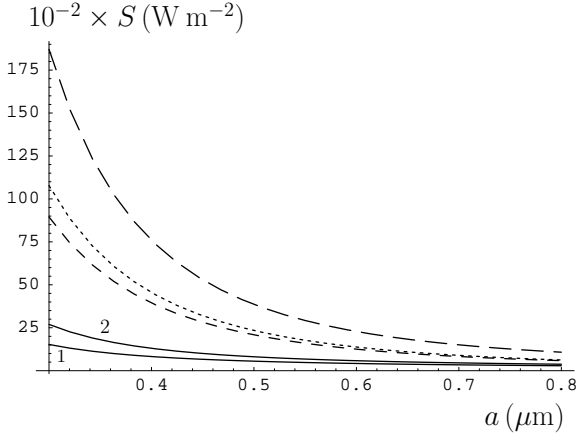}
}
\vspace*{-9cm}
\caption{
Plots of  radiative heat transfer
between two Au plates at
temperatures
$T_1=320$ K and $T_2=300$ K, as a function of separation,
according to
Lifshitz theory for
$\varepsilon=\varepsilon_D$
(short-dashed line),
and to   impedance theory for three
different choices of
impedance:
$Z=Z_N$ (long-dashed line), $Z=Z_D$ (dotted line)
and $Z=Z_t$ (band between
the solid lines 1 and 2).
See text for further
explanation.}
\label{fig:3}
\end{figure}

It should be noted that, according to Eq.(\ref{htim}), the value
of $S$ is
very sensitive to the real part ${\rm Re}\,Z(\omega)$ of
the impedance
function, at angular frequencies $\omega$ of the
order of $k_B
T/\hbar$. It is also important to observe that at
small separations $a$, $S$
receives a  large contribution from the EW.
Since, as seen from Table~1,
thermally exited EW  are of great
importance for the determination of
the thermal correction $\Delta
P(a,T)$ to the Casimir pressure, it is
clear that a measurement of
$S$ would provide an independent verification
of the validity of
the impedance function,  used in the evaluation of
$\Delta
P(a,T)$. We have estimated numerically the heat transfer $S$
for
three choices of the impedance functions,
$Z_D$, $Z_N$ and $Z_t$, that were discussed
in Sec.~6. In Fig.\ 3, we show plots of the radiated power
 for two plates of
Au as a function of the separation
$a$ for $T_1=320$ K and $T_2=300$ K.
The three
lines are for the standard Lifshitz theory with the
Drude
dielectric function in
Eq.~(\ref{PBeq13}) (short-dashed line), for
the
impedance theory with the
impedance $Z_N$ of the normal-skin effect in
Eq.~(\ref{PBeq12})
(long-dashed line), and again for the impedance theory,
but this time
with the impedance $Z_D$ corresponding to the Drude model
in
Eq.~(\ref{PBeq13}) (dotted line). The band between the solid lines
1
and 2 is related to the
impedance function $Z_t$ in Eq.~(\ref{eq4-7}),
with the upper and
lower boundaries corresponding to $\beta=0.08$ eV
and
$\beta=0.125$ eV, respectively. The line for $Z=Z_p$
in
Eq.~(\ref{eq4-5}) is not
displayed, because it reproduces Re$Z_p$ inaccurately
and
the corresponding values for $S$ are over two
orders of magnitude
smaller than those for, say, $Z_t$.
Note that the dielectric permittivity approach with
$\varepsilon$ corresponding to $Z_t$ leads to almost the same results
as are presented by the band between the solid lines 1 and 2 in Fig.~3.
The only difference is that at short separations the lines 1 and 2
are a bit shifted towards smaller values of $S$,
while preserving the same
asymptotic values at large separations.
As we see, Lifshitz theory using the
dielectric permittivity $\varepsilon_D$,
as well as the impedance theory
for $Z=Z_D$
and $Z=Z_N$, both lead to values for $S$ that are several times
larger
than those implied by the impedance $Z_t$ or respective dielectric
permittivity, for separations around
or less than half micrometer. To a large extent, these
large differences are due to considerably
different contributions from the TE EW
in the various models. In Fig.\ 4,
 we show a plot of the relative
contribution to $S$ from TE EW,
for the  models considered in Fig.~3. As we see,
the TE EW play an
important role in the entire range of
separations
considered.

\begin{figure}
\vspace*{-4cm}
\hspace*{-3cm}
\resizebox{0.75\textwidth}{!}{%
 \includegraphics{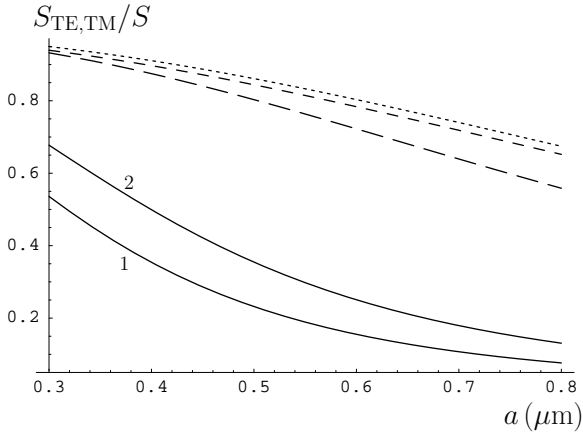}
}
\vspace*{-9cm}
\caption{
Plots of the relative contributions of
TE EW to the total radiative
heat transfer between two Au plates   at
temperatures $T_1=320$ K and
$T_2=300$ K as a function of
separation. Lines are notated as in
Fig.~3.}
\label{plotssigma1}
\end{figure}
Besides heat transfer, another interesting quantity to
consider is
the total emittivity $e(T)$ of the metal, defined
as:
\be
e(T)=\frac{\Phi(T)}{\Phi_{\rm BB}(T)}\;.
\label{bb}
\ee
\noindent
Here,
$\Phi(T)$ is the
total flux of radiation from a unit surface of a
metal
defined in Eq.~(\ref{kirch}),
while $\Phi_{\rm BB}(T)$ is the  flux from a
black
body. According to Stefan law, the latter quantity is equal
to
\be
\Phi_{\rm BB}(T)=\Theta \,T^4\, ,
\label{stefan}
\ee
\noindent
where
$\Theta=5.6704 \times
10^{-8}$ W ${\rm m}^{-2} {\rm K}^{-4}$. For a
polished surface of
gold at $T=295$ K, the tabulated value is
$e\approx0.02$.  In Table~3
we report the calculated values for $e$, corresponding
to the same
models as considered in Fig.~3. The interval of values for
$e$,
appearing in the fifth column in the case of $Z=Z_t$, is
related to the values of $\beta$ considered in
Fig.~3, with the lower and upper
values of $e$ corresponding to
$\beta=0.125$ eV and $\beta=0.08$ eV, respectively.
As is seen in Table~III, the measured emittivity is better described
by the impedance of the normal skin effect and by the generalized
impedance of the infrared optics $Z_t$. However, any definite
conclusion is impossible without information on the precision
of emittivity measurements.

\section{Conclusions and discussion}

In the first part of
this paper we have implemented the thermodynamic test for two
different choices for the
surface impedance of a metal, that are used in the theory of 
the thermal Casimir
force. By making analytic perturbation expansions in powers of small
parameters, we have obtained the asymptotic expressions for the free
energy and entropy of a fluctuating field at low temperatures. This was
done by using the Leontovich impedances of the infrared optics
$Z_i$ and of the Drude model $Z_D$.
The Leontovich impedance of the infrared optics,
extrapolated to low frequencies, withstood the thermodynamic test.
The obtained asymptotic expressions for the free energy and entropy of
a fluctuating field are found to be thermodynamically consistent.
In particular, entropy becomes zero when temperature vanishes, i.e.,
the Nernst heat theorem is satisfied. On the other hand, the Leontovich
impedance of the Drude model was shown to be thermodynamically
inconsistent. In the case of metals with perfect crystal lattices the
entropy at zero temperature was found to be positive and depending on the
parameters of the system in violation of the Nernst heat theorem.

The above conclusions provoke two questions on how to correctly
apply thermal quantum field theory in Matsubara formulation
to real materials. Both the result of this
paper and of \cite{36,37} on the thermodynamic inconsistency
of $Z_D$ and $\varepsilon_D$, respectively, in the theory of the
thermal Casimir force were obtained by using the idealization of perfect
crystal lattice of a metal. In the presence of impurities there is a
nonzero residual relaxation at $T=0$, i.e., $\gamma(0)\neq 0$.
As a result, at very low temperatures the first Matsubara frequencies may
become less than $\gamma(0)$ and the entropy jumps steeply to zero.
For metals with impurities, described by the dielectric permittivity
of the Drude model, the vanishing of entropy at $T=0$ was
demonstrated in \cite{PB26,BrPRE05}. This, however, does not solve
the problem arising for metals with perfect crystal lattices. Such
metals have a nonzero relaxation $\gamma(T)$ at nonzero $T$ and
are commonly used as the basic model in the theory of electron-phonon
interactions. For perfect crystal lattices with no impurities the
Nernst heat theorem is proved in the framework of quantum statistical
physics \cite{PB53}, and the violation of this theorem by the entropy
of a fluctuating field is a problem of great concern. It is our opinion
that the violation of the third law of thermodynamics by the Casimir
entropy calculated using $\varepsilon_D$ and $Z_D$ warns about
the inapplicability of the Drude model in the theory of the thermal
Casimir force.

Another question to discuss is the physical meaning of the zero
Matsubara frequency in the Lifshitz formula, Eq.~(\ref{PBeq3}), or of low
frequencies in the equivalent form of it, Eq.~(\ref{PBeq6}), expressed in terms
of real frequenceis. Should the analytic expression for the impedance of
a metal which is valid for frequencies
around the characteristic frequency $\Omega_c$, be extrapolated
without modifications
to quasistatic frequencies, as we did in Sec.~4, where we dealt with
the impedance of infrared optics? Or, alternatively, should one
use different impedance functions within different frequency regions
in accordance with their applicability conditions? The latter approach
was in fact used in Sec.~5 because the impedance of the Drude model
coincides with the impedance of the infrared optics in the region of
infrared frequencies around $\Omega_c$ and with the impedance of the
normal skin effect in the region of quasistatic frequencies. Our
results demonstrate that in spite of being rather natural
to use different impedance functions in accordance
with the frequency regions of their applicability, and not use any
extrapolation, the actual situation is not so simple. As is shown
in Sec.~4, the extrapolation of the impedance of infrared optics to
zero Matsubara frequency satisfies the thermodynamic test, whereas the
use of the Drude model impedance in Sec.~5, coinciding with the
impedances of the normal skin effect and infrared optics in the
appropriate frequency regions, violates thermodynamics. This should be
compared with the results of \cite{PB31,PB32,PB33,PB34,PB34a} 
devoted to the
Casimir interaction between two dielectrics and between metal and
dielectric. In both cases the account of an actual dielectric response
at quasistatic frequencies (i.e., the account of nonzero dc
conductivity) results in contradiction with thermodynamics, whereas the
extrapolation of the dielectric behavior at high frequencies to zero
frequency satisfies the thermodynamic test. This leads us to
argue that the response function of both a metal and a dielectric
to a real external electromagnetic field of
very low, quasistatic, frequency is not related to the physical
phenomenon of dispersion forces determined by the electromagnetic
fluctuations of high frequencies. In terms of physical processes
occuring at different frequencies the above discussed
extrapolation would imply that in metal bodies the fluctuating
electromagnetic field creates only  pure imaginary currents
related to the
frequency region of infrared optics. However, if the characteristic
frequency belongs to the region of infrared optics, real currents
with typical frequencies of the normal skin effect cannot be
created by the fluctuating field. The deeper understanding of these
gueses may go beyond the scope of the Lifshitz theory.

In the second part of this paper we
have performed  a comparative phenomenological
investigation of the thermal Casimir
force between two parallel metal plates and of
the radiative heat transfer
which occurs between such plates when they
are kept at different
temperatures. Both phenomena are of the same
physical nature because they are
caused by   electromagnetic
fluctuations. As was discussed above, the problem of
the thermal Casimir force meets with difficulties,
and different
controversial appro\-a\-ches to its resolution were proposed
in the
literature. These approaches are based on the use of
various
dielectric functions (the dielectric
permittivities of the Drude and of the
plasma model) or,
alternatively, different forms of the Leontovich
surface impedance.
The selection between the approaches is done by the
comparison
of the obtained results with the requirements of thermodynamics
and
with the experimental data. In this paper we have demonstrated
that
the use of the impedance functions $Z_N(\omega)$ and
$Z_D(\omega)$
(constructed using the dielectric permittivities of the
normal skin effect and of the Drude
model)  to
calculate the thermal correction to the Casimir force
leads to
contradiction with experiment at separations of a few
hundred nanometers.
Recall that earlier the use of the dielectric
permittivity of the Drude
model to calculate the thermal
correction was also shown to be
inconsistent with experiment.
At the same time, both the dielectric permittivity
of the plasma
model and the corresponding impedances related to the
region
of infrared optics are consistent with experiment.

The radiative
heat transfer between   two plates was previously
studied using the
dielectric permittivity of the Drude model
\cite{27,polder,loomis} and the
related impedance $Z_D(\omega)$
\cite{BiPRL}. These approaches,
however, lead to a contradiction with
experiment in the case of the thermal
Casimir force.
Because of this, it is of much interest to investigate the
radiative
heat transfer using the impedance of   infrared optics.
Lack of
reliable experimental data for the power of heat transfer
and resulting uncertainty with the experimental confirmation of
the computations
in \cite{27,BiPRL,polder,loomis} add
importance to this
aim.

Bearing in mind that the radiative heat transfer is very sensitive to
the real part of the impedance function we have constructed
the new
impedance $Z_t$ in the region of infrared optics. The real part
of this
impedance takes into account the tabulated optical data for
the complex index
of refraction extrapolated to low frequencies in
accordance with
general theoretical requirements. The power of heat
transfer calculated with
this impedance is several times less than
previous predictions at
separations of a few hundred nanometers.
These large differences are mainly
explained by different
contributions from the TE EW in the various
models of a metal.

Both physical phenomena of the thermal Casimir force
and of the
radiative heat transfer are finding prospective applications
in
nanotechnology. This makes urgent to carry out new
precise
experiments in order to find what characterization of real metals
is most
adequate for the description of electromagnetic
fluctuations.

\begin{acknowledgement}
{\it Acknowledgements.}
V.B.B. and C.R were partially supported by CNPq
(Brazil) and by FAPESQ-Pb/CNPq (PRONEX).
 G.B. was partially supported by the
PRIN SINTESI.
G.L.K. and V.M.M. were supported by
the
University of Naples Federico II, under the program
"Programma
Internazionale di Modilit$\grave{{\rm a}}$ Docenti e Studenti",
and by the PRIN
SINTESI.
They were also partially supported by CNPq, PRONEX and by
Deutsche Forschungsgemeinschaft grants
436\,RUS\,113/789/0--2 and 0--3.
\end{acknowledgement}


\begin{table}[h]
\caption{\label{tab1} Thermal
correction to the pressure between
Au plates at $T=300\,$K (column 2) and
different contributions to
it as a function of separation computed using the
impedance of the
normal skin effect $Z_N$. See text for further
discussion.}
\begin{center}
\begin{tabular}{ccccccc}
\hline\noalign{\smallskip}
$a\,(\mu\mbox{m})$
& $\Delta P$\,(mPa) & $\frac{\Delta P}{\Delta
P^{\rm IM}}$ &
$\frac{\Delta P_{\rm TE,EW}}{\Delta P}$ &
$\frac{\Delta P_{\rm TE,PW}}{\Delta P}$
& $\frac{\Delta P_{\rm
TM,EW}}{\Delta P} $ &
$\frac{\Delta P_{\rm
TM,PW}}{\Delta P}$ \\
\noalign{\smallskip}\hline\noalign{\smallskip}
0.2 &--12.1 & 5.9 $\times 10^3$ & 0.998 & --9
$\times 10^{-5}$&
2 $\times 10^{-3}$& --2 $\times 10^{-4}$\\
0.25 &--5.4
& 2.6 $\times 10^3$ & 0.997& --7 $\times 10^{-5}$ &
3.6 $\times
10^{-3}$ & --2 $\times 10^{-4}$ \\
0.3 & --2.8 & 1.4 $\times 10^3$ & 0.995&
--3 $\times 10^{-6}$&
5 $\times 10^{-3}$ & --3 $\times 10^{-4}$\\
0.35
&--1.6 & 7.8 $\times 10^2$ & 0.994 & 1.5 $\times 10^{-4}$ &
6 $\times
10^{-3}$ &--3  $\times 10^{-4}$\\
0.4 & --0.96 & 4.7 $\times 10^2$ &
0.992 & 5 $\times 10^{-4}$ &
8 $\times 10^{-3}$&--2 $\times 10^{-4}$ \\
1
&--0.032 & 16 & 0.91 & 0.03 & 0.03 &
0.02\\
\noalign{\smallskip}\hline
\end{tabular}
\end{center}
\end{table}
\begin{table}[h]
\caption{\label{tab2}
Thermal correction to the pressure between
Au plates at $T=300\,$K
(column 2) and different contributions to
it as a function of separation
computed using the impedance of
infrared optics $Z_p$. See text for
further
discussion.}
\begin{center}
\begin{tabular}{ccccccc}
\hline\noalign{\smallskip}
$a\,(\mu\mbox{m})$ & $\Delta P$\,(mPa) & $\frac{\Delta P}{\Delta
P^{\rm IM}}$ &
$\frac{\Delta P_{\rm TE,EW}}{\Delta P}$ &
$\frac{\Delta P_{\rm
TE,PW}}{\Delta P}$ & $\frac{\Delta P_{\rm
TM,EW}}{\Delta P} $ &
$\frac{\Delta
P_{\rm TM,PW}}{\Delta P}$ \\
\noalign{\smallskip}\hline\noalign{\smallskip}
0.2 & --0.0097 & 4.7 & --0.76 &
0.10& 1.55&0.1 \\
0.25 & --0.0081 & 4.0 &--0.43 &0.13  &1.18 &0.12 \\
0.3 &
--0.0070 & 3.4  & --0.26 &0.14  &0.97  &0.14  \\
0.35 & --0.0059 & 2.9
&--0.18 & 0.17& 0.84 &0.17  \\
0.4 & --0.0051 & 2.5 &--0.13 &0.20 &0.73
& 0.20\\
1 & --0.0026 & 1.3 & --0.01& 0.39 & 0.2 &
0.4\\
\noalign{\smallskip}\hline
\end{tabular}
\end{center}
\end{table}
\begin{table}
\caption{\label{tab:table3}
Values   of the emittivity $e$ of Au
at $T=295$
K.}
\begin{center}
\begin{tabular}{cccccc}
\hline\noalign{\smallskip}
Measured & Lifshitz
& $Z=Z_N$    & $Z=Z_D$   &
$Z=Z_t$ & $Z=Z_p$ \\
&($\varepsilon=\varepsilon_D$) & & & & \\
\noalign{\smallskip}\hline\noalign{\smallskip}
0.02  &
0.0098  & 0.023 & 0.0098 & 0.013$\div$0.016$$ &
2.5 $\times 10^{-4}$
\\
\noalign{\smallskip}\hline
\end{tabular}
\end{center}
\end{table}
\end{document}